\documentclass[prd,twocolumn,nofootinbib,showpacs,superscriptaddress]{revtex4-1}

\usepackage{amsfonts}
\usepackage{amsmath}
\usepackage{amssymb}
\usepackage{bm}
\usepackage{dcolumn}
\usepackage{epsfig}
\usepackage{graphicx}
\usepackage{graphics}
\usepackage[latin1]{inputenc}
\usepackage{latexsym}
\usepackage{rotating}
\usepackage{hyperref}
\usepackage[usenames]{color}
\usepackage{float}
\usepackage{xspace} 
\usepackage{mathrsfs}
\usepackage{subfigure}
\usepackage{multirow}

\newcommand{\mc}{\ensuremath{{\cal{M}}}}

\def\prd{{\it Phys. Rev.} D~}

\def\apjl{{\it Astrophys. J.} Lett~}
\def\apj{{\it Astrophys. J.}}
\def\Msun{M_\odot}
\def\PRD{{\it Phys. Rev.} D~}

\def\mnras{{\it MNRAS~}} 

\def\aap{{\it A\&A~}}
\def\na{{\it New Astronomy~}}
\def\ptp{{\it Progress of Theoretical Physics~}}

\def\lesssim{\mathrel{\hbox{\rlap{\hbox{\lower4pt\hbox{$\sim$}}}\hbox{$<$}}}}
\def\gtrsim{\mathrel{\hbox{\rlap{\hbox{\lower4pt\hbox{$\sim$}}}\hbox{$>$}}}}
\def\alt{\mathrel{\hbox{\rlap{\hbox{\lower4pt\hbox{$\sim$}}}\hbox{$<$}}}}
\def\agt{\mathrel{\hbox{\rlap{\hbox{\lower4pt\hbox{$\sim$}}}\hbox{$>$}}}}
\def\apjs{{\it ApJS}}
\def\araa{{\it ARA\&A}}
\def\ssr{{\it Space Sci. Rev.}}

\def\gta{\ifmmode {\mathbin{\lower 3pt\hbox   
    {$\,\rlap{\raise 5pt\hbox{$\char'076$}}\mathchar"7218\,$}}}
    \else {${\mathbin{\lower 3pt\hbox
    {$\rlap{\raise 5pt\hbox{$\char'076$}}\mathchar"7218\,$}}}
    $}\fi}
\def\lta{\ifmmode {\,\mathbin{\lower 3pt\hbox   
    {$\,\rlap{\raise 5pt\hbox{$\char'074$}}\mathchar"7218\,$}}}
    \else {${\mathbin{\lower 3pt\hbox
    {$\rlap{\raise 5pt\hbox{$\char'074$}}\mathchar"7218\,$}}}
    $}\fi}

\newcommand{\beq}{\begin{equation}}
\newcommand{\eeq}{\end{equation}}
\newcommand{\bea}{\begin{eqnarray}}
\newcommand{\eea}{\end{eqnarray}}

\newcommand{\WVU}{\affiliation{Department of Physics and Astronomy, West Virginia University, White Hall, Morgantown, WV 26506, USA}}
\newcommand{\CU}{\affiliation{Institute of Astronomy, Madingley Road, CB3 0HA Cambridge, United Kingdom}}
\newcommand{\CA}{\affiliation{Jet Propulsion Laboratory, California Institute of Technology, 4800 Oak Grove Drive, Pasadena, CA 91106, USA}}
\newcommand{\NCSA}{\affiliation{NCSA, University of Illinois at Urbana-Champaign, Illinois 61801, USA}}
\newcommand{\ED}{\affiliation{School of Mathematics, University of Edinburg, King's Building, Edinburgh, EH9 3JZ, United Kingdom}}

\begin{document}

\title{Detection of eccentric supermassive black hole binaries with pulsar timing arrays: Signal-to-noise ratio calculations}
\author{E. A. Huerta}
\email{elihu@illinois.edu}
\WVU\NCSA%
\author{Sean T. McWilliams}
\WVU%
\author{Jonathan R. Gair}
\CU\ED%
\author{Stephen R. Taylor}
\CA%


\begin{abstract}
We present a detailed analysis of the expected signal-to-noise ratios of supermassive black hole binaries on eccentric orbits observed by pulsar timing arrays. We derive several analytical relations that extend the results of Peters and Mathews~\cite{peters} to quantify the impact of eccentricity in the detection of single resolvable binaries in the pulsar timing array band. We present ready-to-use expressions to compute the increase/loss in signal-to-noise ratio of eccentric single resolvable sources whose dominant harmonic is located in the low/high frequency sensitivity regime of pulsar timing arrays. Building upon the work of Phinney~\cite{Phinney:2001} and Enoki and Nagashima~\cite{Eno:2007}, we present an analytical framework that enables the construction of rapid spectra for a stochastic gravitational wave background generated by a cosmological population of eccentric sources. We confirm previous findings which indicate that, relative to a population of quasi-circular binaries, the strain of a stochastic, isotropic gravitational wave background generated by a cosmological population of eccentric binaries will be suppressed in the frequency band of pulsar timing arrays. We quantify this effect in terms of  signal to noise ratios in a pulsar timing array.
\end{abstract}

\maketitle

\section{Introduction}

It is believed that supermassive black holes (SMBHs) with masses between \(10^6 \Msun -10^9\Msun\) are ubiquitous in galactic nuclei~\cite{Begelman:1980,Kormendy:1995,Ferrarese:2005}. According to the accepted framework of hierarchical structure formation, massive galaxies are formed by continuous accretion of gas from cosmic web filaments or through galactic mergers~\cite{Blanco:1978MNRAS,Kauffmann:2000MNRAS}. This latter mechanism naturally leads to the formation of SMBH binaries in the merged galaxy remnants. As the SMBHs sink in the potential well of the remnant galaxy due to dynamical friction, stars within the binary orbit are quickly ejected. An  SMBH merger can only take place if additional mechanisms operate to remove energy and angular momentum from the binary, e.g., friction from a spherical Bondi accretion flow~\cite{Escala:2004ApJ}, a circumnuclear  gas disk~\cite{Roedig:2011MNRAS}, slingshot scattering of stars on low angular momentum orbits intersecting the binary~\cite{frank,Quinlan:1996NewA,YU:2002MNRAS}, etc. If any of these mechanisms can drive the orbit to sufficiently small separations, gravitational wave (GW) emission can take over and drive the binary system the rest of the way to coalescence within a Hubble time~\cite{Peters:1964,peters,Th300,multipole,KI,preto}. 

Regarding the orbital properties of SMBH binaries, scattering interactions between individual stars and SMBH binaries can potentially drive the binaries to large orbital eccentricities, particularly when the binaries retain significant eccentricities at the end of the dynamical friction phase~\cite{Quinlan:1996NewA,Sesana:2006ApJ,Berentzen:2009ApJ,KhanF:2012ApJ}, whereas SMBH binaries embedded in sufficiently massive prograde self-gravitating gas disks may acquire eccentricities as large as \(e\sim 0.6-0.8\) by the time gravitational radiation takes over the dynamical evolution of the system~\cite{Roedig:2011MNRAS}. Furthermore, SMBH binaries embedded in counter rotating disks may be driven to very large values of eccentricity \(e\sim 1\)~\cite{Nixon:2011,Schnittman:2015x}, even though the binary can flip and realign with the disk~\cite{Roeding:2014MNRAS}.

The gravitational radiation emitted during the inspiral of binaries with masses \(10^{6}\Msun - 10^{9}\Msun\) out to redshifts \(z\lesssim 1\) will be detectable by Pulsar Timing Arrays (PTAs)~\cite{Zhu:2014MNRAS,Arzoumanian:2014ApJ,NANOMaura:2013,Hobbs:2013CQG,KramerCha:2013CQG,Manchester:2013CQG,Moore:2014,Lazio:2013CQG,Yardley:2010,IPTA:2010}. PTAs are capable of detecting cosmic string networks, primordial GWs, an unresolved stochastic GW background generated by a large population of compact binary sources~\cite{Siemens:2013,Haasteren:2011,Haasteren:2009,Anholm:2009,KocSas:2011,Ses:2008} and GWs from individual binary systems~\cite{SesanaVec:2010CQG,SesanaVecVol:2009MNRAS}. 

Given the significant attention that eccentric compact binaries have attracted as potential sources of GWs  and electromagnetic radiation~\cite{MayerL:2013CQG,Huerta:2014,GairL:2013,PAmaro:2012CQG,Huerta:2013a}, there is a need to study the effect of eccentricity both in terms of source detection and parameter estimation for individually resolvable sources, and for the detection of a stochastic GW background in the context of PTAs. Our understanding on the effect of eccentricity on potential GW sources for PTAs has gradually improved from the seminal work of Quinlan~\cite{Quinlan:1996NewA}, and recent theoretical and numerical studies that have shed light on the impact of eccentricity and environmental effects in suppressing the low frequency GW background in the PTA band~\cite{Sesana:2013CQG,HaimanZ:2009,Sesana:2011M,KocsisH:2012,KocsisB:2012ApJ,Ravi:2014,Taylor:2015,Schnittman:2015}.

In this article we build upon the work of Phinney~\cite{Phinney:2001} and Enoki and Nagashima~\cite{Eno:2007} by constructing an analytical framework that enables the construction of rapid spectra for a stochastic GW background generated by a population of eccentric sources. We then employ a prescription for  the evolution of the BH mass function taken from \cite{McST:2014}, and combine it with our results  to compute the signal-to-noise ratios (SNRs) of a stochastic GW background generated by a population of eccentric binaries, with the SNR in that case being derived from a cross-correlation statistic, given that matched filtering cannot be applied to a stochastic signal. We also derive several analytical summations that expand upon the results of Peters and Mathews~\cite{peters} (PM hereafter) to explore in detail the effect of eccentricity on the GW strain and the matched-filter SNRs of individually resolvable sources.

Our studies show conclusively that the SNR of eccentric binaries is non-negligibly attenuated for eccentricity values \((e\gtrsim 0.7)\). However, binaries with low to moderate values of eccentricity  \((0\lesssim e\lesssim 0.6)\) will have SNRs comparable to their quasi-circular counterparts. This suggests, in principle, that the detection of a population of eccentric binaries may be possible and would provide new insights on the formation channels of SMBH binaries and their cosmological evolution. However, it is still necessary to show that the imprints  of eccentricity can be accurately extracted from GW observations with PTAs. We defer the study of this important issue to future work. 

This article is organized as follows: in Section~\ref{gn_s} we provide a succinct description of the properties of eccentric binary systems and derive analytical relations that are of importance for eccentric SMBH binaries observed by PTAs. In Section~\ref{spectrum} we provide analytical results for the energy density and the characteristic amplitude of the GW spectrum, and discuss at length the effect of eccentricity on these two observables. In Section~\ref{sec:snr_rs} we apply this calculated strain to compute SNRs for both single resolvable sources and a stochastic population of eccentric binaries with \(e\in[0,\,0.9]\). We summarize our findings and describe future directions for research in Section~\ref{sec:conclusions}. Throughout this article we use geometric units with \(G=c=1\).


\section{Power from individual eccentric binaries}
\label{gn_s}

Consider a binary system with component masses \((m_1, \, m_2)\), such that \(m_1> m_2\), \(M=m_1+m_2\), and whose orbital rest-frame frequency is given by \(f_{\rm orb}= \omega/2\pi\). If the system evolves from an initial state with nonnegligible eccentricity \(e\) and semi-major axis \(a\), then the binary radiates GWs in the whole spectrum of harmonics. Furthermore, as shown by PM~\cite{peters}, the relative power \(P(n)\) radiated in the $n$'th harmonic is given by:
\begin{equation} 
P(n) = \frac{32}{5}\frac{m_1^2 m_2^2 \left(m_1+m_2\right)}{a^5} g(n,e)\,,
\label{power}
\end{equation}
\noindent where

\begin{eqnarray}
\label{gsform}
g(n,e) &=& \frac{n^4}{32} \Bigg[  \bigg\{J_{n-2}(ne) -2e J_{n-1}(ne) \\\nonumber&+& \frac{2}{n} J_{n}(ne) + 2eJ_{n+1}(ne)-J_{n+2}(ne)\bigg\}^2\\\nonumber&+&\left(1-e^2\right)\bigg\{ J_{n-2}(ne) -2J_{n}(ne)+J_{n+2}(ne)\bigg\}^2 \\\nonumber&+& \frac{4}{3n^2}J^2_{n}(ne)\Bigg]\,. 
\end{eqnarray}

\noindent Using Bessel's equation and recurrence relations, one can re-write Eq.~\eqref{gsform} as follows: 

\begin{eqnarray}
\label{newgsform}
g(n,e) &=& \frac{n^4}{32}\Bigg[   \frac{J^2_n}{n^2}\left(  2-\frac{4}{e^2}\right)^2 +  J^{'2}_n \left(  \frac{4}{e} -4e\right)^2 \\\nonumber&+& \frac{2 J_n J'_n}{n}\left(  2-\frac{4}{e^2}\right)\left(  \frac{4}{e} -4e\right) \\\nonumber&+& J^2_n\left(1-e^2\right)\left(  \frac{4}{e^2} -4\right)^2 + \frac{J_n^{'2}}{n^2}\left(1-e^2\right)\left(\frac{4}{e}\right)^2 \\\nonumber&-& \frac{2 J_n J'_n}{n} \frac{4\left(1-e^2\right)}{e} \left( \frac{4}{e^2}-4\right) + \frac{4}{3n^2}J_n^2\Bigg]\,.
\end{eqnarray}

\noindent Note that Eq.~\eqref{newgsform} corrects a typo in Eq. (A1) of PM \cite{peters}. As shown in PM:

\begin{equation}
F(e)=\sum_{n=1}^{\infty}  g(n,e) =\frac{1 + \frac{73}{24}e^2+ \frac{37}{96}e^4}{\left(1-e^2\right)^{7/2}}\,.
\label{eval_g}
\end{equation}
\noindent Hence, averaging  over one period of the elliptical motion, the average rate at which the binary system radiates energy is given by:
\begin{eqnarray}
\langle P \rangle &=& \sum_{n=1}^{\infty} P(n)\,,\nonumber\\ 
&=&  \frac{32}{5}\frac{m_1^2 m_2^2 \,M}{a^5 \left(1-e^2\right)^{7/2}} \left(1 + \frac{73}{24} e^2+ \frac{37}{96}e^4\right)\,.
\label{average_power}
\end{eqnarray}
Another interesting quantity that involves the object \(g(n\,,e)\) is the GW strain root-mean-square (rms) amplitude. As discussed in~\cite{thorne}, the rms amplitude and the energy radiated in the $n$'th harmonic are related through:
\begin{equation}
h_n = \frac{1+z}{\pi d_L}\frac{\sqrt{\dot{E}_n}}{ n\, f_{\rm orb}}\,,
\label{rms}
\end{equation}
\noindent where \(z\) is the redshift. Since the luminosity, \(\dot{E}\), emitted by the system averaged over one complete orbit is given by
\begin{equation}
\dot{E} = \frac{32}{5}{\mc^{10/3}}\left(2\pi f_{\rm orb}\right)^{10/3}\sum_{n=1}^{\infty}g(n,\,e)\,,
\label{lum}
\end{equation}
\noindent then Eq.~\eqref{rms} can be re-written as follows:
\begin{equation}
h_n = 2\sqrt{\frac{32}{5}}\frac{\mc^{5/3}}{n d_L} \left(2\pi f_{\rm orb}\right)^{2/3}\sqrt{g(n,\,e)}(1+z)\,,
\label{rms_new}
\end{equation}
\noindent where \(\mc= M\,\eta^{3/5}\) is the chirp mass, and \(\eta = m_1 m_2/M^2\) represents the symmetric mass ratio. It is possible to obtain a similar expression to the average power by considering the quantity 
\begin{equation}
\sum_{n=1}^{\infty} h^2_n = \frac{32}{5}\frac{\mc^{10/3}}{d^2} \left(2\pi f_{\rm orb}\right)^{4/3}\sum_{n=1}^{\infty}\frac{g(n,\,e)}{\left(n/2\right)^2}\,,
\label{pow_rms}
\end{equation}
\noindent where \(d= d_L/(1+z)\). This quantity has heretofore been evaluated numerically using a given number of harmonics to ensure a specified accuracy. However, one can derive an exact closed form for the sum appearing on the right-hand side of this expression, as shown in Appendix~\ref{Apen_PN}:  
\begin{equation}
H(e)=\sum_{n=1}^{\infty}\frac{g(n,\,e)}{n^2} = \frac{4-\sqrt{1-e^2} }{12\sqrt{1-e^2}}\,.
\label{governsq}
\end{equation}
\noindent Thus, Eq.~\eqref{pow_rms} takes the simple form:
\begin{equation}
\sum_{n=1}^{\infty} h^2_n = \frac{32}{15}\frac{\left(4-\sqrt{1-e^2}\right) }{\sqrt{1-e^2}}\frac{\mc^{10/3}}{d^2} \left(2\pi f_{\rm orb}\right)^{4/3}\,.
\label{sum_rms}
\end{equation}
\noindent In the following Section, we will use a similar approach to derive new analytical relations to explore the signatures that a population of eccentric binaries may imprint on a stochastic background of gravitational radiation and on single resolvable sources. 


\section{Stochastic background of a population of eccentric binaries}
\label{spectrum}
Following Ref.~\cite{Phinney:2001}, one can define the total GW energy density per logarithmic frequency interval observed today from a population of (instantaneously monochromatic) sources as:
\begin{eqnarray}
\label{gwener}
{\cal E}_{\rm GW} &\equiv& \int_0^\infty \rho_c \Omega_{\rm GW}(f) \frac{{\rm d}f}{f} \equiv \int_0^\infty \frac{\pi}{4}  f^2 h_c^2(f) \frac{{\rm d}f}{f} \nonumber\\ &=&  \int_0^\infty  \int_0^\infty N(z) \frac{1}{1+z}  f_r \frac{{\rm d}E_{\rm GW}}{{\rm d}f_r} {\rm d}z \frac{{\rm d}f}{f} .
\end{eqnarray}
\noindent The rate of mergers per unit comoving volume which occur between redshift \(z + \mathrm{d}z\) is given by \(N(z) \mathrm{d}z\). Furthermore, \(\rho_c\) represents the rest-mass energy that would be required to close the Universe~\cite{Phinney:2001}
\begin{equation}
\rho_c = \frac{3 H^2_0}{8\pi}\,.
\label{crit_den}
\end{equation}
In practice, we can replace \(N(z) \mathrm{d}z\)  by a differential rate and integrate over source parameters, but for the moment we shall assume that the population is composed of identical sources. If the sources have eccentricity then they will no longer be instantaneously monochromatic. Instead, we can regard the emission at each harmonic to represent a separate population of sources. Based on this observation, and following Enoki and Nagashima~\cite{Eno:2007}, we have that:

\begin{eqnarray}
{\cal E}_{\rm GW} &=& \sum_{n=1}^\infty {\cal E}_{{\rm GW},n}, \qquad \mathrm{with}\nonumber\\
{\cal E}_{{\rm GW},n} &=& \int_0^\infty  \int_0^\infty N(z) \frac{1}{1+z}  f_{n,r} \frac{{\rm d}E_{{\rm GW},n}}{{\rm d}f_{n,r}} {\rm d}z \frac{{\rm d}f_{n}}{f_n}\,,\nonumber\\
\end{eqnarray}
\noindent  where $f_n$ represents the frequency of the $n$'th harmonic observed today, and $f_{n,r}=(1+z) f_n$ is the frequency of the harmonic in the rest frame. The amount of energy radiated in GWs into the $n$'th harmonic as the frequency of the $n$'th harmonic changes from $f_{n,r}$ to $f_{n,r}+{\rm d}f_{n,r}$ is given by:
\begin{equation}
\label{rad_ener}
\frac{{\rm d}E_{{\rm GW},n}}{{\rm d}f_{n,r}} {\rm d}f_{n,r}\,.
\end{equation}
\noindent This outgoing energy is measured in the source's rest frame, and is integrated over the entire radiating lifetime of the source and over all solid angles~\cite{Phinney:2001}. Using the relations
\begin{equation}
\label{pre_eq}
4\pi^2\,f^2_{\rm orb}\, a^3 = M \quad {\rm and} \quad  \omega = 2\pi \,f_{\rm orb}\,,
\end{equation}
\noindent with Eq.~\eqref{lum}, we find that: 
\begin{eqnarray}
\label{en_1}
\frac{{\rm d}E_{{\rm GW},n}}{{\rm d}f_{n,r}} &=& \frac{{\rm d}E_{{\rm GW},n}}{{\rm d}t_r} \frac{{\rm d}t_r}{{\rm d}f_{n,r}}\,, \\
\label{en_2}
\frac{{\rm d}E_{{\rm GW},n}}{{\rm d}t_r} &=& \frac{32}{5} \left({\cal M}\,\omega\right)^{10/3}\, g(n,e)\,, \\
\label{en_3}
\frac{{\rm d}f_{n,r}}{{\rm d}t_r} &=& n \frac{{\rm d}f_{\rm orb}}{{\rm d}t_r} =  \frac{96}{5} \frac{n\, F(e)}{2\pi}  {\cal M}^{5/3} \omega^{11/3}\,,
\end{eqnarray}
\noindent where \(F(e)\) was defined in Eq.~\eqref{eval_g}. Combining these, we find
\begin{eqnarray}
\frac{{\rm d}E_{{\rm GW},n}}{{\rm d}f_{n,r}} &=& \frac{\left(2\pi\right)^{2/3} {\cal M}^{5/3}}{3  F(e)}  \frac{g(n,e)}{n} f_{\rm orb}^{-\frac{1}{3}} \nonumber\\
&=& \frac{\pi^{2/3} {\cal M}^{5/3}}{3 F(e)\left(1+z\right)^{1/3}} \frac{g(n,e)}{\left(n/2\right)^{2/3}} f^{-\frac{1}{3}}\,.
\label{rad_reac}
\end{eqnarray}
The energy density in the background per logarithmic frequency interval is then given by
\begin{eqnarray}
\label{endes}
&&\rho_c \Omega_{\rm GW} (f) = \frac{\pi}{4}  f^2 h_c^2(f)  =\\&&\frac{{\cal M}^{5/3} \, \left(\pi f\right)^{2/3}}{3} \int_0^\infty\sum_{n=1}^\infty \frac{1}{F(e)}\frac{g(n,e)}{\left(n/2\right)^{2/3}}  \frac{N(z)}{\left(1+z\right)^{1/3}} {\rm d} z.\nonumber
\end{eqnarray}
\noindent Note that in the quasi-circular limit \((n=2, \,e\rightarrow 0)\), Eq.~\eqref{endes} recovers the results presented in Ref.~\cite{Phinney:2001}. 


\subsection{Estimating the number of merger events in unit comoving volume \(N(z)\)}
\label{subtle}
One important ingredient in the calculation of the stochastic spectrum, energy density and, ultimately, the SNR with which a population of GW sources can be detected is the number of mergers that occur between redshift \(z\) and \(z+\rm{d}z\), i.e., \(N(z)\). For the systems under consideration, i.e., binaries with total masses between \(10^{6-9}\Msun\), our knowledge of the numbers and mass distributions of SMBHs has changed considerably with the advent of large scale surveys~\cite{McConnellN:2013,KomatsuE:2011} and recent theoretical studies~\cite{McST:2014,SesanaSystematic:2013,LinO:2010}. However, deriving a robust model for the computation of \(N(z)\) is a complex problem due to the large uncertainties inherent in several aspects of the calculation, e.g., the poorly constrained rate of BH migration toward the center of merging galaxies caused by interactions with dark matter, gas, and stars; the possibility of multiple BH interactions in the event that the BH migration is inefficient, etc.~\cite{Farris:2014,Farr:2014,Tanaka:2013,McKernan:2013,KocsisH:2012,TanakaMenou:2012,Lippai:2009,HaimanZ:2009,Rov:2014,VanW:2014,Bonoli:2014,Fiaconni:2013,Chap:2013,KhanF:2012}.  With these caveats in mind, we use the estimate for \(N(z)\) described in Ref.~\cite{McST:2014}, which we will review here for completeness. 
 
We need to estimate the comoving number density of BHs with masses between \(M_{\bullet}\) and \(M_{\bullet}+\textrm{d}M_{\bullet}\).  BH masses are strongly correlated with the bulge masses of their hosts and so this is equivalent to considering the distribution of galaxy bulge masses. The starting point for such an estimate is an
empirical model known as the Schechter function~\cite{Schechter:1976}, given by 
\begin{equation}
\phi(M)\textrm{d}M = \varphi \,M^{\alpha} \exp\left(-M\right)\textrm{d}M \,,
\label{schter}
\end{equation}
\noindent where $\varphi$ and \(\alpha\) represent the normalization of the luminosity function and the faint-end slope parameter, respectively. The Schechter function is a power law that is truncated at large masses.  For the most massive galaxies of interest, we need to amend this function to account for the observed excess of mass in the brightest cluster galaxies and other very massive elliptical galaxies.
Following Ref.~\cite{LinO:2010}, we do this by adding a Gaussian component to Eq.~\eqref{schter}:
\begin{eqnarray}
\phi(M)\textrm{d}M &=& \left(\varphi + \varphi_{\rm massive}\right) \textrm{d}M  = \varphi \,M^{\alpha} \exp\left(-M\right)\textrm{d}M \nonumber\\&+&  \hat{\phi}\, \exp\left(-\frac{1}{2}\left(\frac{2.5 \log M}{\sigma}\right)^2 -1\right)\textrm{d}M\,,\nonumber\\
\label{schter_imp}
\end{eqnarray}
\noindent where \(\hat{\phi}\) and \( \varphi\)  are normalization factors to describe the brightest cluster galaxies and less massive galaxies, respectively.  
We try to encapsulate in a conservative way the current knowledge we have from galaxies that host BHs with masses \(\sim 10^9 \Msun\) such as M87. Hence, following Ref.~\cite{McST:2014} we set  \(\hat{\phi} = \varphi\) and \(\sigma=0.58\), which ensures at least one M87-mass source in our sample. The comoving density of BHs can be constructed from the Schechter function by replacing
\begin{equation}
M\rightarrow \frac{M_{\bullet}}{\mathsf{M}}\,\quad {\rm{with}} \quad \mathsf{M} = \frac{1.2\times10^8}{1+z}\Msun\,.
\label{map_of_mass}
\end{equation}
\noindent Here \( M_{\bullet} \) denotes the BH mass and \( \mathsf{M}\) is a Schechter parameter that represents the characteristic mass at the turnover of the mass function. This particular prescription is consistent with observational data~\cite{vanDokkum:2010}. We set the normalization of the luminosity function to have the constant value 
\begin{figure}[htp]
\centerline{
\includegraphics[height=0.35\textwidth,  clip]{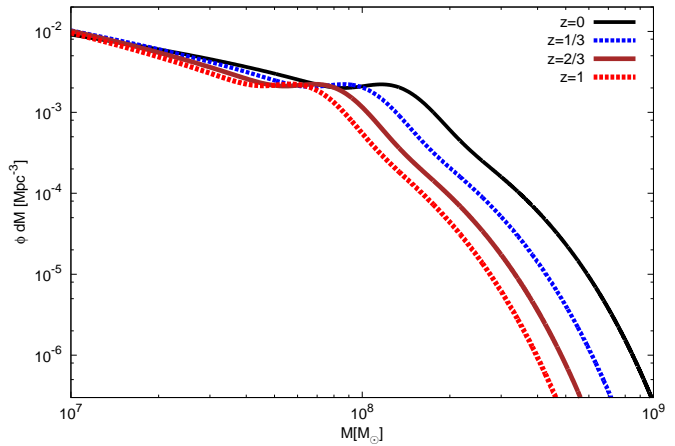}
}
\caption{Redshift evolution of the black hole mass function given by Eq.~\eqref{schter_imp}.}
\label{sch_plot}
\end{figure}
\begin{equation}
\varphi \equiv 3\times 10^{-3} \,{\rm Mpc}^{-3}\,.
\label{phi_cst}
\end{equation}
\noindent This choice is in good agreement with results presented in Ref.~\cite{ColeS:2001} at low redshifts and using the cosmological parameters presented in Ref.~\cite{KomatsuE:2011}. Observational data suggests that \(\varphi\) might have a mild dependence on redshift. However, following Ref.~\cite{McST:2014}, we ignore the redshift dependence of \(\varphi\) because it is a small effect that has a negligible influence on the total GW signal.  Finally, ensuring that the redshift dependence of the faint-end slope parameter \(\alpha\) satisfies mass conservation, one finds that~\cite{McST:2014}:
\begin{equation} 
\alpha\approx -2+ \frac{0.52}{1+z}\,.
\label{alpha_param}
\end{equation}
\noindent We can reconstruct the BH mass function introduced in Ref.~\cite{McST:2014} by plugging Eqs.~\eqref{map_of_mass}--\eqref{alpha_param} into Eq.~\eqref{schter_imp} --- see Figure~\ref{sch_plot}. This approach reproduces the results presented in Ref.~\cite{SesanaSystematic:2013} at a \(2\sigma\) level. Following Ref.~\cite{McST:2014}, we express the number density of mergers \(N(z)\) by assuming that it is proportional to the product of the number density of the constituent black holes, as shown in Eq. (8) of Ref.~\cite{McST:2014}. Using this approach, we evaluate the integral
\begin{equation} 
N_0 =  \int_{z_{\rm min}}^{z_{\rm max}} \frac{N(z)}{\left(1+z\right)^{1/3}}  {\rm d} z\,,
\label{eval_int}
\end{equation}
\noindent where \(z_{\rm min}=0\) and \(z_{\rm max}=1\). Assuming that all systems in the Universe have the same eccentricity we find that 
\begin{equation}
N_0 =\left\{\begin{array}{cl}
 2.63\times10^{-3}\,{\rm Mpc}^{-3}\,, & 7\lesssim \log M \lesssim 7.9\,,\\
  1.16\times10^{-3}\,{\rm Mpc}^{-3}\,, & \log M\gtrsim 7.9\,.\end{array}\right.
\label{mer_rates}
\end{equation}
We have used the mass ranges quoted above motivated by the fact that mass function has a break around \(\log M \sim 7.9\) for all redshifts of interest. Note that even though this is a rough approximation, we have verified that this choice does not have a strong influence on the results.


\subsection{Ready to use expressions for the gravitational wave energy density and the characteristic amplitude of the gravitational wave spectrum}
\label{sec:ready_to_use}
Having derived all of the ingredients to compute the GW energy density and the characteristic amplitude of the GW spectrum, and using the most recent results for the cosmological parameters released by the Plank Collaboration in Ref.~\cite{Planck:2014} to compute the critical density of the Universe defined by Eq.~(\ref{crit_den}), we can derive ready-to-use expressions for the energy density and the characteristic amplitude of the GW background. We carry out this calculation in two steps. We first provide a pedagogic example in which the eccentricity of the SMBH binary population is assumed to be constant. Thereafter, we address the likely physical scenario in which the eccentricity evolves as a function of frequency due to GW emission.

\subsubsection{Compact binary population with fixed eccentricity}
\label{fixed_e_evolution}
Assuming that the eccentricity of the binary population is fixed, we can derive an analytical expression that reproduces the sum in Eq.~\eqref{endes} to better than \(0.01\%\) in the eccentricity range \(e_0\in[0,\,0.95]\) (see Appendix~\ref{fundamental_sum}), namely: 
\begin{equation}
A(e_0)=\sum_{n=1}^\infty \frac{g(n,e_0)}{\left(n/2\right)^{2/3}}=  \frac{1 + \frac{1467}{1024} e_0^2 - \frac{115}{12288}e_0^4 + \frac{227}{32768}e_0^6}{\left(1-e_0^2\right)^{5/2}}\,.   
\label{gn23}
\end{equation}   
\noindent For later convenience, let us define the function:
\begin{equation}
\label{bfunc}
B(e_0) \equiv \frac{A(e_0)}{F(e_0)}= \left(1-e_0^2\right) \frac{\left(1 + \frac{1467}{1024} e_0^2 - \frac{115}{12288}e_0^4 + \frac{227}{32768}e_0^6\right)}{1 + \frac{73}{24}e_0^2+ \frac{37}{96}e_0^4}\,.
\end{equation}
\noindent Using Eq.~\eqref{bfunc}, the energy density and the characteristic amplitude of the GW background take the form:

\begin{eqnarray}
\label{ensim_new}
 \Omega_{\rm GW} (f) &=& 3.6\times10^{-10} \left(\frac{{\cal M}}{10^8 \, \Msun}\right)^{5/3}   \left(\frac{f}{1\,{\rm yr}^{-1}}\right)^{2/3}\nonumber\\&\times&\left( \frac{N_0 }{10^{-3} \, {\rm Mpc}^{-3}}\right)   B(e_0)\,,\\
 \label{stch_new}
h_c(f) &=& 5.0\times10^{-16} \left(\frac{{\cal M}}{10^8 \, \Msun}\right)^{5/6} \left(\frac{f}{1\,{\rm yr}^{-1}}\right)^{-2/3}\nonumber\\&\times&\left( \frac{N_0 }{10^{-3} \, {\rm Mpc}^{-3}}\right)^{1/2}  \sqrt{B(e_0)}\,.
\end{eqnarray}

\begin{figure}[ht!]
\centerline{
\includegraphics[height=0.35\textwidth,  clip]{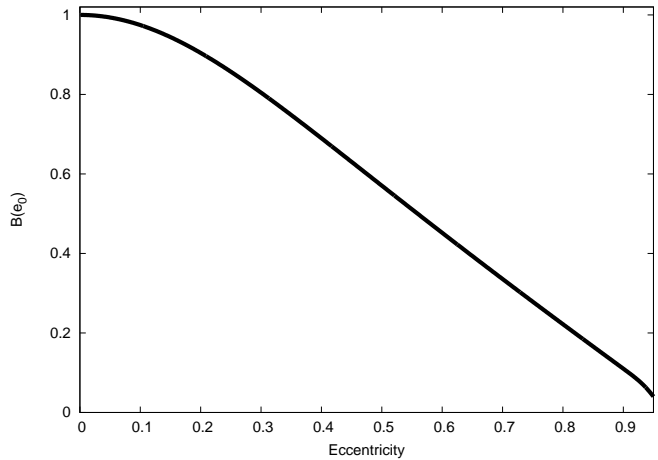}
}
\caption{Attenuation factor, \(B(e_0)\), as defined in Eq.~\eqref{bfunc}, which describes the decrease in the emitted energy density for a population of compact sources with fixed eccentricity. In light of Eqs.~\eqref{ensim_new} and~\eqref{stch_new},  the present-day energy density \( \Omega_{\rm GW} (f) \) is maximized for a population of quasi-circular compact binaries, whereas its value is decreased by a factor \(\sim 10\) for a population of highly eccentric systems \((e_0\sim0.9)\). Similarly, the characteristic amplitude of the GW spectrum steadily decreases as the eccentricity of the compact binary population increases.}
\label{becorr}
\end{figure}

\noindent In Figure~\ref{becorr} we plot the attenuation function \(B(e_0)\) (see Eq.~\eqref{bfunc}). We notice that both the energy density and the characteristic amplitude of the GW background are maximized for a population of quasi-circular binaries, and steadily decrease for increasing values of eccentricity. These results give the energy density and typical strain of a GW background generated by binaries with fixed eccentricity and chirp mass.

\subsubsection{Compact binary population with evolving eccentricity}
\label{evolving_e_evolution}
To describe a compact binary population whose eccentricity is evolving, we notice that for a given initial eccentricity \(e_0\) at a fiducial initial orbital frequency \(f_0\), the eccentricity depends only on the orbital frequency: \(e=e(f_{\rm orb}, \, e_0)\). Each harmonic \(n\) contributes to the signal at an observed frequency \(f= n\, f_{\rm orb}/(1+z)\). Hence, including the frequency evolution of the eccentricity entails replacing the argument of the \(g(n, e),\, F(e)\) functions in Eq.~\eqref{endes} by 
\begin{equation}
\label{map}
e(f_{\rm orb};\, e_0) = e\left( \frac{1+z}{n}f;\, e_0\right)\,.
\end{equation}

\begin{figure}[ht]
\centerline{
\includegraphics[height=0.35\textwidth,  clip]{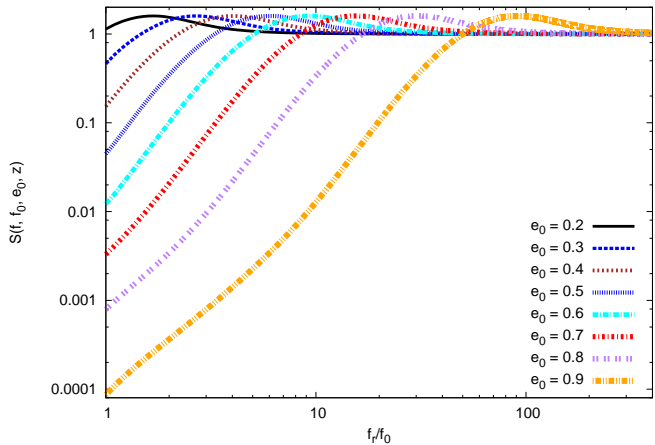}
}
\caption{The panel shows the frequency evolution of the function \(S(f, f_0, e_0, z)\), given by Eq.~\eqref{l_sum} for several values of initial eccentricity \(e_0\). The \(x\)-axis shows the ratio \(f_r/f_0\), where \(f_r = \left(1+z\right)f\).}
\label{s_function}
\end{figure}

\noindent We shall use the dictionary \(e\rightarrow e(f_{\rm orb})\) given by Eq. (3.12) of Ref.~\cite{Yunes:2009}, which is robust for \(e\in[0,\,0.9]\), namely:
\begin{equation}
e(f_{\rm orb};\, e_0)  \rightarrow \frac{16.83 - 3.814\, \beta^{0.3858}}{16.04 + 8.1\,\beta^{1.637} }\,,
\label{new_map}
\end{equation}
\noindent where \(\beta = \chi^{2/3}/\sigma_0\) and \(\chi=  f_{\rm orb}/f_0\), with \(f_{\rm orb} =  \left(1+z\right)f/n\), and 

\begin{equation}
\label{sigma_zero}
\sigma_0 = \frac{e_0^{12/19}}{1-e_0^2}\left(1+\frac{121}{304}e_0^2\right)^{870/2299}\,.
\end{equation} 

\noindent We substitute Eq.~\eqref{new_map} into Eq.~\eqref{endes} to obtain the function: 
\begin{equation} 
\label{l_sum}
S(f, f_0, e_0, z)=\sum_{n=1}^\infty \frac{1}{F(e(f_{\rm orb};\, e_0))}\frac{g(n,e(f_{\rm orb};\, e_0))}{\left(n/2\right)^{2/3}}\,.
\end{equation}
\noindent In Figure~\ref{s_function} we show the frequency evolution of the function \(S(f, f_0, e_0, z)\) for several values of initial eccentricity \(e_0\). It is worth pointing out that these results are in excellent agreement with Ref.~\cite{Eno:2007}, even though we have used a different approach to parameterize the orbital frequency evolution. We have found several interesting properties of the generating function  \(S(f, f_0, e_0, z)\):
\begin{itemize}
\item The location of the maxima follows a simple relation given by:
\begin{equation}
x^{\rm max} \cong \frac{1293}{181}\left(  \frac{e^{12/19}_0}{1-e^2_0}\Bigg[ 1+\frac{121}{304}e^2_0\Bigg]^{870/2299}\right)^{3/2}\,,
\label{maxima}
\end{equation}
\noindent where \(x=f_r/f_0\).
\item The maxima of the \(S(f, f_0, e_0, z)\) function is the same for all values of \(e_0\) and is given by
\begin{equation}
S(f, f_0, e_0, z)^{\rm max} = \frac{373}{234}\,.
\label{maxima_of_s}
\end{equation}
\item Two additional properties that  \(S(f, f_0, e_0, z)\) must satisfy are:
\subitem  \(S(f, f_0, e_0=0, z)\equiv 1\)\,,
\subitem  \(\lim_{f\to\infty} S(f, f_0, e_0, z) \rightarrow 1\)\,.
\end{itemize}
In light of this analysis, we have constructed a function that has these generic properties. We found it convenient to split the function in two pieces given its distinct properties before and after it reaches $S=1$. The points at which \( S(f, f_0, e_0, z) \equiv 1\) are given by:
\begin{equation}
x^{\rm fixed} \cong \frac{3620\, e_0}{841\,\left(1-e_0^2\right)^3}\left(1-\frac{370}{243}e_0^2 + \frac{132}{269}e_0^4\right)\,.
\label{fixed_points}
\end{equation}
\noindent In the domain \(x\gtrsim x^{\rm fixed}\), we propose the following ansatz for  \(S(f, f_0, e_0, z)\):
\begin{equation} 
\label{the_eq}
S_{\rm high}(f, f_0, e_0, z)=1 + a(e_0)[x-b(e_0)]e^{-c(e_0)\,x}\,,
\end{equation}
\noindent where \(x= f_r/f_0\) and the eccentricity dependent coefficients \(a(e_0)\), \(b(e_0)\), \(c(e_0)\) are given by: 
\begin{eqnarray}
\label{b_coe}
b(e_0)&=& x^{\rm fixed}\,,\\
\label{c_coe}
c(e_0) &=& \frac{1}{x^{\rm max}-x^{\rm fixed}}\,,\\
\label{a_coe}
a(e_0)&=& \frac{S^{\rm max}-1}{x^{\rm max}-x^{\rm fixed}}\,\exp\left(c(e_0)\, x^{\rm max}\right)\,.
\end{eqnarray}
\noindent It is worth pointing out that for low values of eccentricity (\(e_0\lesssim 0.2\)), Eq.~\eqref{the_eq} reproduces the main features of  \( S(f, f_0, e_0, z)\) throughout the domain \(x\geq 1\). When we consider \(e_0\gtrsim 0.2\), we need to replace the low frequency evolution using the following relation 
\begin{equation}
\label{s_low_x}
S_{\rm low}(f, f_0, e_0, z)=d(e_0)\,x^{\left(29-s(e_0)\right)/7}\,e^{-g(e_0)\,x} \,,
\end{equation}
\noindent where the coefficients \((d(e_0),\, s(e_0),\, g(e_0))\) are determined by enforcing that \(S_{\rm low}\) has the correct value at \(x=1\) and \(x=x^{\rm fixed}\), and that  \(S'_{\rm low}(x^{\rm fixed})=S'_{\rm high}(x^{\rm fixed})\). The transition from $S_{\rm low}$ to $S_{\rm high}$ is at the point $x^{\rm fixed}$. 

\begin{figure}[ht]
\centerline{
\includegraphics[height=0.35\textwidth,  clip]{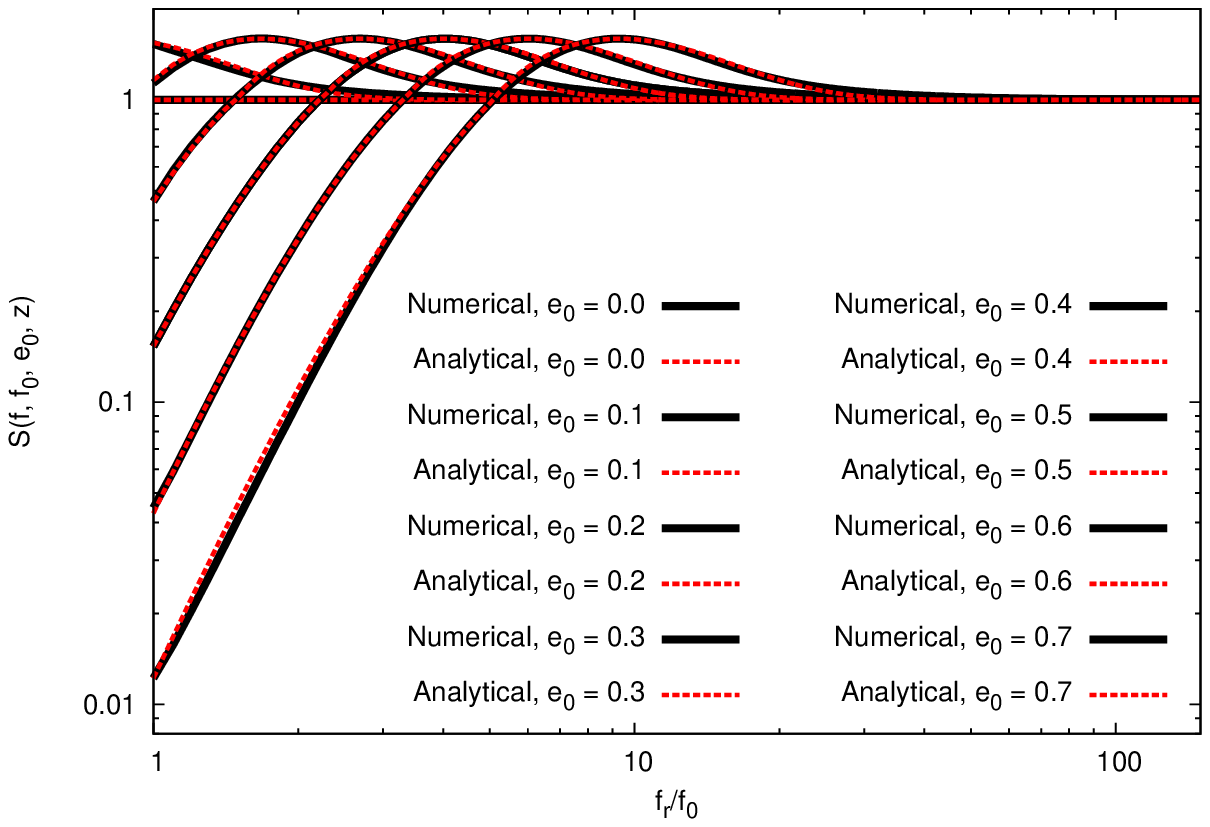}
}
\centerline{
\includegraphics[height=0.35\textwidth,  clip]{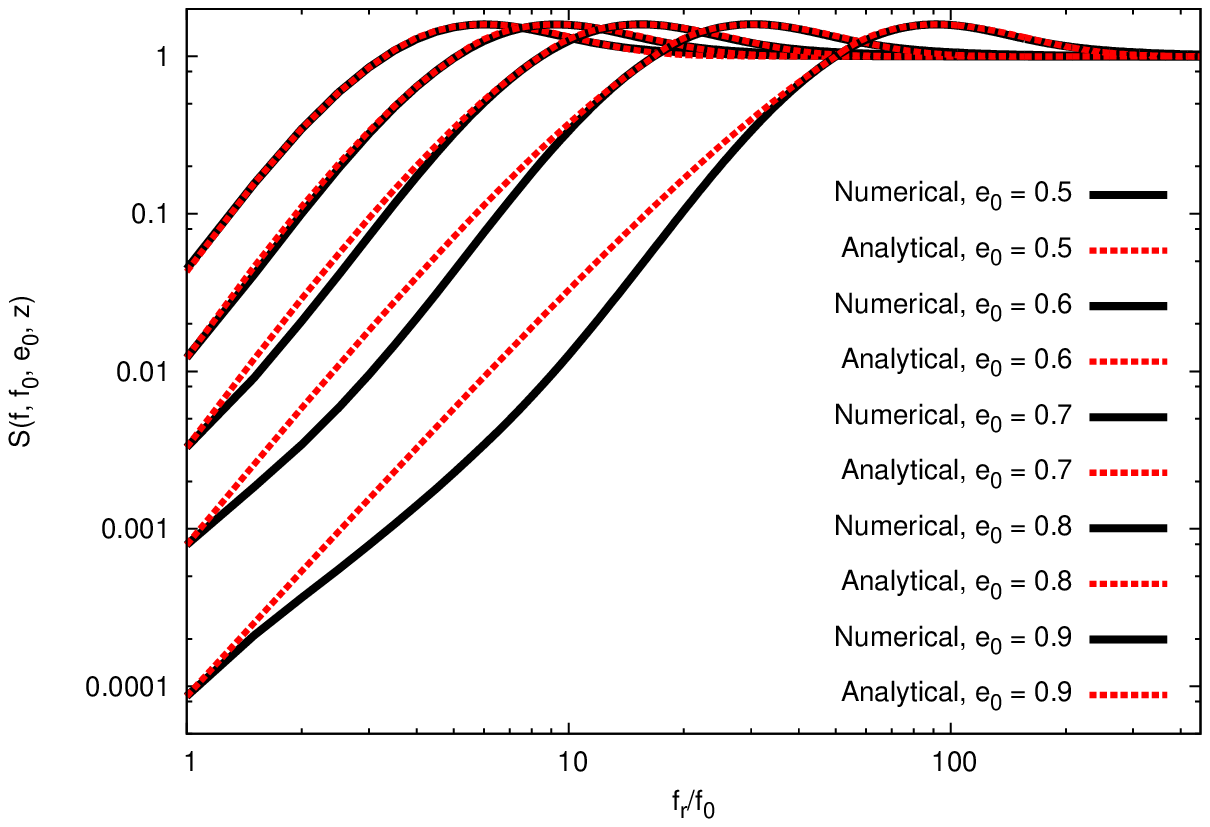}
}
\caption{The top panel shows a direct comparison between the numerical solution of the sum \(S(f, f_0, e_0, z)\) given by Eq.~\eqref{l_sum} and the analytical solution we have constructed using Eqs.~\eqref{the_eq} and~\eqref{s_low_x}. This analytical solution reproduces the full numerical solution for systems with eccentricity \(e_0\lesssim 0.7\) with an accuracy better than \(10\%\) in the domain \(x\geq 1\). The largest deviation occurs for populations with \(e_0= 0.7\), which start to deviate from self-similar solutions in the low frequency regime (\(1\lesssim x \lesssim x^{\rm fixed}\)). Bottom panel: populations  with higher eccentricity have a non self-similar evolution in the domain  \(1\lesssim x \lesssim x^{\rm fixed}\), but we can still provide an approximate description in this regime using Eq.~\eqref{s_low_x}. Please note that Eq.~\eqref{the_eq} provides a reliable description of the spectra for any value of eccentricity \(e_0\in[0,\,0.9]\) for \(x \gtrsim x^{\rm fixed}\), and Eq.~\eqref{s_low_x} smoothly asymptotes to the numerical solution when \(x\rightarrow x^{\rm fixed}\).}
\label{s_function_a}
\end{figure}

We have found that \(S_{\rm high}(f, f_0, e_0, z)\), given by Eq.~\eqref{the_eq}, can accurately describe the full numerical solution of Eq.~\eqref{l_sum} for \(e_0\in[0,\, 0.9]\) in the domain \(x\gtrsim x^{\rm fixed}\). This is possible because the numerical solution has self-similarity properties that are captured by Eqs.~\eqref{maxima}-~\eqref{a_coe}. We have attempted to provide a similar parameterization for the spectra in the domain \(1\lesssim x \lesssim x^{\rm fixed}\) and have found that self-similarity is present for populations with \(e_0\lesssim 0.7\). Populations with larger eccentricities have two properties that deviate from self-similarity in the domain \(1\lesssim x \lesssim x^{\rm fixed}\): (a) the slope of the spectra evolves as a function of eccentricity; (b) the spectra develops a bulging at lower frequencies that becomes more pronounced for increasing values of eccentricity. These two properties are clearly shown in the bottom panel of Figure~\ref{s_function_a}. The parameterization we propose in Eq.~\eqref{s_low_x} captures the evolution of the spectra as a function of  eccentricity with the parameter \(s(e_0)\). Using both Eqs.~\eqref{the_eq} and~\eqref{s_low_x}, we can analytically reproduce the numerical solution of Eq.~\eqref{l_sum}  for systems with eccentricity \(e_0\lesssim 0.7\) with an accuracy better than \(10\%\) in the domain \(x\geq 1\) --- note that the largest deviation between the numerical and analytical solutions occurs for populations with \(e_0= 0.7\). The discrepancy arises because, even if we have captured the evolution of the slope of the spectra as a function of eccentricity, the numerical solution presents an additional bulging at low frequencies that is not equally present in all spectra. Indeed, the bottom panel of Figure~\ref{s_function_a} shows that this feature becomes increasingly pronounced for highly eccentric populations in the low frequency domain. However, we notice that \(S_{\rm low}(f, f_0, e_0, z)\) still provides an approximate description of the spectra in the low frequency domain that smoothly asymptotes to the numerical solution when \(x\rightarrow x^{\rm fixed}\). This is an important property, since this is the region where the signal is most likely to be detected. Therefore, given the ever-increasing attenuation of the spectra for very large eccentricities, it seems that the analytical framework we have constructed covers the entire domain of detectable stochastic signals. Finally, we note that, by construction, our analytical approach satisfies \(S(f, f_0, e_0=0, z)\equiv 1\) and   \(\lim_{f\to\infty} S(f, f_0, e_0, z) \rightarrow 1\). In future studies that aim at modeling SMBH binaries that evolve in stellar environments or embedded in counter rotating disks that may drive the eccentricity to large values \(e_0\sim1\), it will be necessary to modify the framework described above by including a non-self-similar evolution for the low frequency evolution part of the spectrum, in particular for eccentricities \(e_0\gtrsim0.7\).

The analytical approximation to the spectra of eccentric populations we have constructed above provides a robust description of the imprint of eccentricity over a wide range of parameter space. Given its simplicity, it provides an ideal tool to be implemented in detection pipelines. We utilize this result in the following Section to compute the expected signal-to-noise ratio with which a cosmological population of SMBH binaries with nonnegligible eccentricity can be detected with PTAs.

\section{Signal-to-noise ratios for pulsar timing arrays}
\label{sec:snr_rs}

In this Section we discuss in detail the prospects of detecting a cosmological population of inspiralling  SMBH binaries with PTAs. Current studies suggest that the expected signal from these events may comprise a superposition of two distinct contributions: (i) a stochastic background generated by the incoherent superposition of gravitational radiation emitted from the whole SMBH population~\cite{Rajagopal:1995,Jaffe:2003,Ses:2008}; and (ii) GW signals that stand above the background and can be individually resolved~\cite{Ses:2009,KocSas:2011}. The motivation to consider these two complementary cases stems from the fact that an inhomogenous combination of multiple sources emitting in the same frequency bin can adopt several configurations in the timing residuals, such as a nearly isotropic distribution over the sky or a few bright spots in the sky if they superpose coherently~\cite{Ses:2008}. There has been a vigorous research program to develop data analysis techniques in the limiting cases of an isotropic stochastic background which, as described in the previous Section, may be described by a power law spectrum~\cite{Jenet:2005,JenHob:2006,Anholm:2009,Haasteren:2009,Haasteren:2011,Siemens:2013}, for single monochromatic GW sources~\cite{Jenet:2004,Sesana:2010,Corbin:2010,Yardley:2010,LeeKJ:2011} and, more recently, for anisotropic GW backgrounds~\cite{2013PhRvD..88f2005M,2013PhRvD..88h4001T,2014PhRvD..90h2001G}, although we will not discuss these further here.

\subsection{Sensitivity of PTAs to single resolvable sources and a stochastic gravitational wave background}
\label{sec:sensitivity_sin_and_sto}
The sensitivity curves of PTAs to continuous waves and a stochastic GW background have been discussed at length in Ref.~\cite{Thrane:2013,Moore:2014}. If we define \(\sigma_{\rm rms}\) as the rms timing noise, and \(1/ \Delta t\) as the cadence of the measurements, then combining Eqs.~(40) and (42) of Ref.~\cite{Thrane:2013}, the dimensionless effective noise amplitude for the timing residuals induced for a stochastic GW background is given by: 
\begin{equation}
h^2_N(f)= f S_n(f) =  24\pi^2 \Delta t\, \sigma^2_{\rm rms}\, f^{3}\,.
\label{ena}
\end{equation}
\noindent Assuming a total observation time \(T_{\rm obs}\),  the analysis presented in~\cite{Moore:2014} shows that the power law integrated  sensitivity curve for a PTA's response to a stochastic GW background  has a sharp cut-off in sensitivity at a frequency \(f= T_{\rm obs}^{-1}\). On the other hand, for individually resolvable sources, the maximum sensitivity is attained around frequencies  \( T_{\rm obs}^{-1}\), and there is a slow diminishing in sensitivity below this value. Assuming a quadratic timing model, Ref.~\cite{Moore:2014} shows that the dimensionless effective noise amplitude for continuous waves can be modeled as a two-part power law in \(f\).  This two-part power law, as given in Ref.~\cite{Moore:2014}, is formally a continuous sum of the two components, but it will prove convenient for us to approximate it as a piecewise combination of the components, namely:
\begin{eqnarray}
\label{high_sens_cw}
&&h_{c,\, \rm{high}}(f) ={\cal{B}} \,f^{\frac{3}{2}}\,, \quad {\rm for}\, f\gtrsim \frac{2}{T_{\rm obs}}\,,\\
\label{low_sens_cw}
&&h_{c,\, \rm{low}}(f) = {\cal{C}}\,f^{-\frac{3}{2}} \,, \quad {\rm with}\\
\label{cst_high}
&&{\cal{B}} =\left(\frac{36}{N_p\left(N_p -1\right)}\right)^{1/2}\,\sqrt{\Delta t}\,\sigma_{\rm rms}\,, \\
\label{cst_low}
&&{\cal{C}} = \frac{8\,{\cal{B}}}{T^3_{\rm obs}}  .
\end{eqnarray}

\noindent The quantity $h_c(f)$ is the characteristic strain of noise fluctuations in the detector, which is required to compute the SNR using Eq.~(\ref{sumofsnr}) below. We note that the transition frequency value \(f_{\rm trans}= 2\,T^{-1}_{\rm obs}\) at which \( h_{c,\, \rm{high}}(f) =h_{c,\, \rm{low}}(f)\) is simply an approximate value for which the two-part power law representation of the total sensitivity reproduces a fully numerical Bayesian analysis~\cite{Moore:2014}. 

We emphasize that the effective sensitivities above differ depending on the detection statistic being assumed, and this has occasionally resulted in some confusion when calculating sensitivity curves, particularly their spectral slopes, throughout the literature.  We have assumed in Eq.~\eqref{ena} that the stochastic background is searched for using a cross-correlation statistic, whereas in Eqs.~\eqref{high_sens_cw}--\eqref{cst_low}, we assume that continuous-wave sources are searched for using matched filtering. 

Having described the prescription we will use for the sensitivity of PTAs to detect continuous wave sources and a stochastic GW background, we will now compute the expected SNR of single resolvable sources.


\subsection{Signal-to-noise ratio calculations for single resolvable sources}
\label{sec:single}

Several recent studies have explored the ability of PTAs to resolve GW sources individually. For instance, assuming the existence of a population of quasi-circular monochromatic sources, an array of pulsars which are equally-sampled every two weeks for ten years, and making several other simplifications regarding the nature of the data sets, Ref.~\cite{BabakandSes:2012} concluded that \(N_s\)  sufficiently loud sources with \({\rm SNRs}\gtrsim 10\) can be resolved and localized in the sky with a network of \(3N_s\) pulsars. Building upon this study, Ref.~\cite{Pet:2013} demonstrated that it was possible to: (i) recover the SNR of injected signals to within a few percent; (ii) infer the sky localization to within a few degrees; and (iii) resolve the frequency at which the signals were injected to better than 0.1 nHz. To put this latter result in context, a PTA that collects data for a time span of \(T_{\rm obs}\) cannot in principle distinguish two GW frequencies separated by less than \(\Delta f\sim 1/T_{\rm obs} \sim 3\)nHz for \(T_{\rm obs}=10\)yr. If the algorithm introduced in Ref.~\cite{Pet:2013} is capable of determining sub-Fourier bin precision to the level of 0.1 nHz, this means that they are capable of resolving up to 30 sources per frequency bin.

A more conservative approach to estimate the number of GW sources that can be individually resolved with a PTA was presented in Ref.~\cite{Boyle:2012}. Basic counting arguments suggest that a PTA with \(N_p\) pulsars can characterize up to \(2N_p/7\) chirping GW point sources per GW frequency bin or \(2N_p/6\) monochromatic sources. This is just the number of measurements (an amplitude and a phase per pulsar) divided by the number of parameters characterizing a single GW source (7 for a chirping binary and 6 for a monochromatic binary). We therefore expect a PTA to be sensitivity limited when every GW frequency bin has more than \(2N_p/7\) sources. At present there are more than 20 pulsars in the IPTA with rms timing residuals \(\sigma_{\rm rms} < 1 \mu {\rm s}\), and a few pulsars with \(\sigma_{\rm rms} < 100 {\rm ns}\)~\cite{Dem:2009}. With the advent of the Chinese five hundred meter spherical aperture telescope~\cite{Smits:2009} and the Square Kilometer Array (SKA)~\cite{Lazio:2009}, there will be a major leap in sensitivity. A conservative estimate suggests that the SKA could detect more than twenty thousand pulsars, including hundreds of them with \(\sigma_{\rm rms}\) that will match or supersede the best pulsars currently known. Such a PTA may no longer be a detector capable only of detecting a stochastic GW background (i.e., a confusion-limited detector) but may become a point source telescope capable of carrying out matched-filtering GW searches ~\cite{Boyle:2012}.  In view of this bright prospect, we now compute the SNRs of eccentric sources in the frequency band of PTAs.

Since binaries on eccentric orbits radiate in a wide spectrum of harmonics \(n\) of the mean orbital frequency, we can write the SNR as:

\begin{eqnarray}
\label{sumofsnr}
\rho_{\ell}^2(n, f_{\rm orb}) &=& \frac{h_{cw}^2(n, f_{\rm orb})}{h_{c,l}^2(n f_{\rm orb})}\,,
\end{eqnarray}
\noindent with \(\ell = [\rm{low}, \, {\rm high}]\) and~\cite{PauTrip:2010}:

\begin{equation}
h_{cw}^2(n, f_{\rm orb})  = \frac{h^2_n\,n f_{\rm orb}\, T_{\rm obs}}{1+z}\,,
\label{compact_form_single}
\end{equation}

\noindent where \(h_n\) is given by Eq.~\eqref{rms_new}. Eq.~\eqref{compact_form_single} can be interpreted as the averaged squared amplitude multiplied by the number of cycles completed during the observation time \(T_{\rm obs}\). In general, the total SNR of a single resolvable source can be written as:

\begin{equation}
\rho^2 \equiv \sum_{n=1}^{n_{\rm max}} \rho_{{\rm low}}^2(n, f_{\rm orb}) + \sum_{n_{\rm max}+1}^{\infty}\rho_{{\rm high}}^2(n, f_{\rm orb})\,,
\label{snr_I}
\end{equation}
\noindent where $n_{\rm max}$ is given by \(n_{\rm max}\, f_{\rm orb} =  f_{\rm trans}\).  Now, bearing in mind that the sensitivity for continuous wave sources is given by a piecewise function, let us consider the low frequency component. Using Eq.~\eqref{low_sens_cw} we find that:
\begin{eqnarray}
\label{low_snr_part}
\rho^2_{\rm{low}} &=&{\hat{\cal{C}}}\,\sum_{n=1}^{n_{\rm max}}\, \left(\frac{n}{2}\right)^{2}\,g(n,e_0)\, f_{\rm orb}^{16/3}\,,\\
\label{low_snr_cst_int}
{\hat{\cal{C}}}&=& \frac{4\,\,\sqrt[3]{2}\,\pi^{4/3}\,N_p \left(N_p-1\right)}{45}\frac{ T^7_{\rm obs}\, {\cal M}^{10/3}}{ d_L^2 \left(1+z\right)^2\Delta t\, \sigma^2_{\rm rms}}\,,\nonumber\\
\label{low_part}
\end{eqnarray}
\noindent where we have used \(f=n\,f_{\rm orb}/(1+z)\) in the last line. In the case where most of the detectable signal is contained in modes with \(n<n_{\rm max}\), we can use an analytical form for the sum  in Eq.~\eqref{low_snr_part}. In Appendix~\ref{Apen_PN} we show that:
\begin{eqnarray}
\label{first_odd_low_fre}
G(e_0) &=& \sum_{n=1}^{\infty} \left(\frac{n}{2}\right)^{2}\ g(n,\,e_0) = \frac{1}{\left(1-e_0^2\right)^{13/2}}\Bigg[ 1 + \frac{85}{6}e_0^2 \nonumber\\&+&  \frac{5171}{192}e_0^4 +\frac{1751}{192} e_0^6 + \frac{297}{1024}e_0^{8}\Bigg]\,.
\end{eqnarray}
\noindent Hence, summing over all the harmonics enables us to recast Eq.~\eqref{low_part} as follows:
\begin{equation}
\rho^2_{\rm{low}} =  {\hat{\cal{C}}}\,G(e_0)\, f_{\rm orb}^{16/3}\,.
\label{low_part_tot}
\end{equation}
\noindent We can find a similar expression for the high frequency contribution,  namely:
\begin{eqnarray}
\label{high_snr_part}
&&\rho^2_{\rm{high}}= {\hat{\cal{B}}}\,\sum_{n=n_{\rm max}}^{\infty}\,\frac{g(n,e_0)}{\left(n/2\right)^4}\, f_{\rm orb}^{-2/3}\,,\\
\label{high_snr_cst_int}
&&{\hat{\cal{B}}}= \frac{4\,\,\sqrt[3]{2}\,\pi^{4/3}\,N_p \left(N_p-1\right)}{45}\frac{ T_{\rm obs}\, {\cal M}^{10/3}\left(1+z\right)^4}{ d_L^2 \Delta t\, \sigma^2_{\rm rms}}\,.\nonumber\\
\end{eqnarray}
\noindent If the first harmonic \(n=1\) is located within the high frequency regime (\(\gtrsim f_{\rm trans}\)), then no detectable signal occurs in the low frequency regime, so \(n_{\rm max}=1\) and we can analytically evaluate the sum in Eq.~\eqref{high_snr_part}. In Appendix~\ref{Apen_PN} we show that this sum  is given by

\begin{equation}
\label{second_odd_low_fre}
Y(e_0) =  \sum_{n=1}^{\infty}\frac{g(n,\,e_0)}{(n/2)^4} = 1 - \frac{1}{3}e_0^2\,,
\end{equation}

\noindent and the high frequency contribution can be expressed as:

\begin{equation}
\rho^2_{\rm{high}} =  {\hat{\cal{B}}}\,Y(e_0)\, f_{\rm orb}^{-2/3}\,.
\label{high_part_lit}
\end{equation}

\noindent We can re-write the low and high frequency contributions to the SNR in a convenient way using the transformation \( u=f_{\rm orb}/f_{\rm trans}\):

\begin{equation}
\rho^2 =\left\{\begin{array}{cl}
  \left(1+z\right)^{-2}{\cal{L}}\, G(e_0)\, u^{16/3}\,, \quad u\ll 1\,,\\
  \left(1+z\right)^{4}{\cal{L}}\,  Y(e_0)\, u^{-2/3}\,,\quad u\geq 1\,, \end{array}\right.
\label{total_SNR}
\end{equation}

\noindent where:

\begin{eqnarray}
\label{cst_uno}
{\cal{L}} &=& \frac{4\,\,\sqrt[3]{2}\,\pi^{4/3}\,N_p \left(N_p-1\right)}{45}\frac{ T^{5/3}_{\rm obs}\, {\cal M}^{10/3}}{ d_L^2 \Delta t\, \sigma^2_{\rm rms}}\,.
\end{eqnarray}

\noindent We note that the requirement that $u\ll 1$ in the first part of Eq.~\eqref{total_SNR} is due to the fact that eccentric sources will emit in a wide range of harmonics.  For more moderate eccentricities, this requirement is weakened, such that Eq.~\eqref{total_SNR} applies to all orbital frequencies in the limit of very small eccentricity.

To give a sense of scale, we can reexpress Eq.~\eqref{total_SNR} as:

\begin{eqnarray}
\label{ready_SNR}
\rho^2 &=&\hat{\rho}^2\left\{\begin{array}{cl}
\left(1+z\right)^{-2}\,G(e_0)\, u^{16/3}\,, u\ll 1\,,\\
\left(1+z\right)^{4}\,Y(e_0)\, u^{-2/3}\,,\,\, \,\, u\geq 1\,,\end{array}\right.\\
\label{prefact}
\hat{\rho}^2_{\rm}&=& 4.26\times10^{-2} \,N_p\left(N_p-1\right)\left(\frac{{\cal M}}{10^8\, \Msun}\right)^{10/3}   \nonumber\\&\times& \left(\frac{T_{\rm obs}}{10\,{\rm yr}}\right)^{5/3}\left( \frac{100\, {\rm Mpc}}{d_L}\right)^2 \nonumber\\&\times&\left(\frac{100\,{\rm ns}}{\sigma_{\rm rms} }\right)^2  \left(\frac{0.05\,{\rm yr}}{\Delta t}\right)\,.
\end{eqnarray}

\begin{figure}[ht]
\centerline{
\includegraphics[height=0.35\textwidth,  clip]{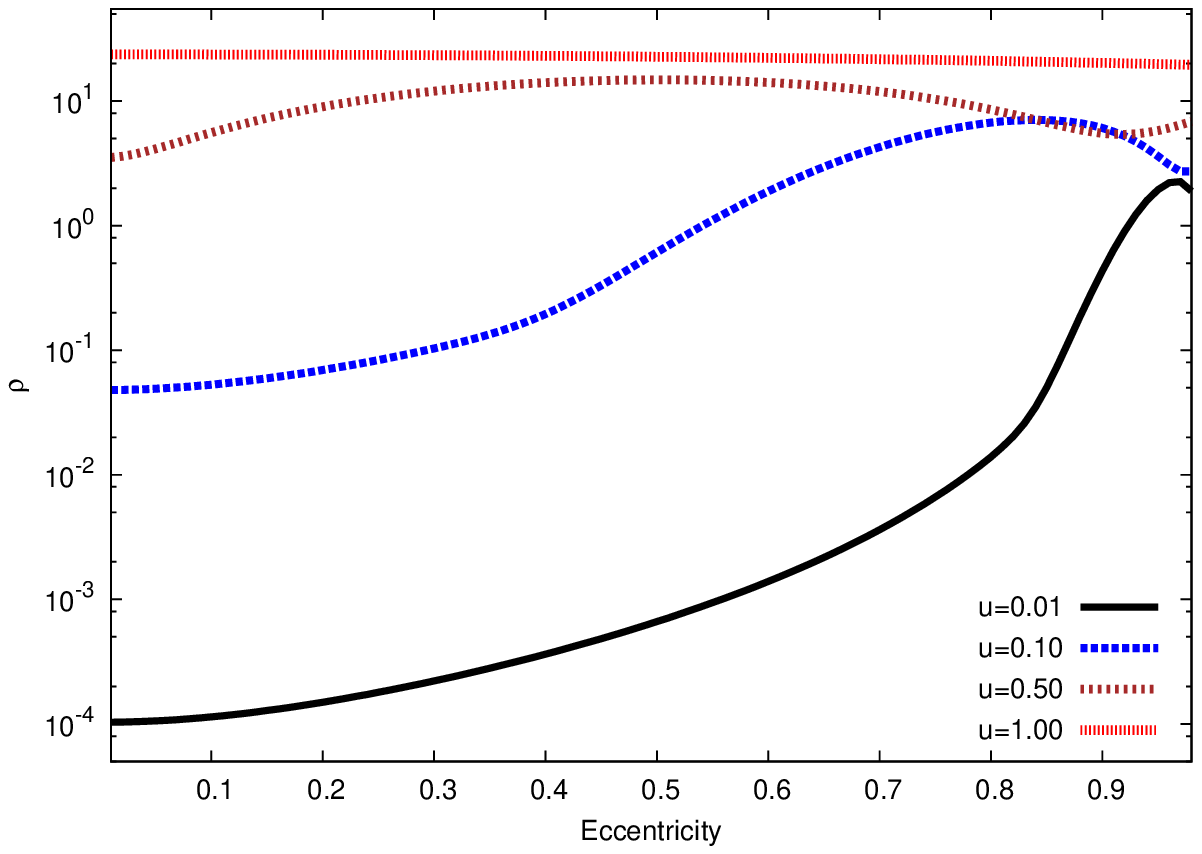}
}
\centerline{
\includegraphics[height=0.35\textwidth,  clip]{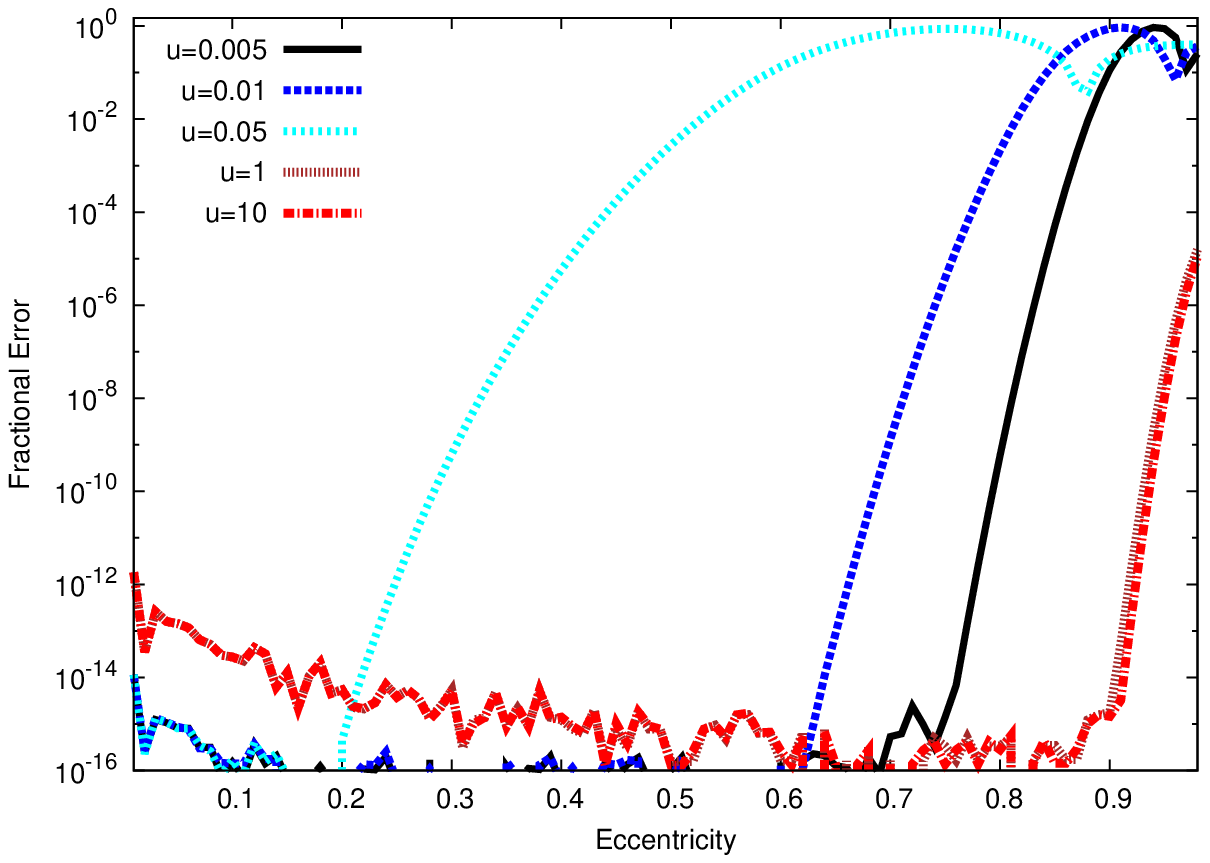}
}
\caption{Expected signal-to-noise ratio \(\rho\) for sources that may be detected in the frequency band of PTAs assuming \(N_p=10\), \(d_L =100\,{\rm Mpc}\), \(z=0.022\), \( \sigma_{\rm rms} = 100\,{\rm ns}\), \( \Delta t =  0.05\,{\rm yr}\) and \(M=10^9 \Msun\) (see Eq.~\eqref{ready_SNR_transition}).  The top panel shows the enhancement in  \(\rho\) at low frequencies (\(u \ll 1\)), and the corresponding attenuation at higher frequencies. We also compare the performance of the expressions given in Eqs.~\eqref{low_part_tot} and~\eqref{high_part_lit} with the actual numerical evaluation of Eq.~\eqref{ready_SNR_transition}.}
\label{single_snr}
\end{figure}

\noindent Finally, in the case that individual sources are emitting in the transition regime between low and high frequency sensitivity (i.e., $f_{\rm orb} < f_{\rm trans}$, but the eccentricity is large enough that significant signal is contained in harmonics with $nf_{\rm orb} > f_{\rm trans}$), the total SNR is given by:

\begin{eqnarray}
\rho^2 &=&\hat{\rho}^2_{\rm}\Bigg[\frac{1}{\left(1+z\right)^{2}}\,\sum_{n=1}^{n_{\rm max}}\, \left(\frac{n}{2}\right)^{2}g(n,e_0)\, u^{16/3} \nonumber\\&+& \left(1+z\right)^{4}\,\sum_{n_{\rm max}+1}^{N_{\rm max}}\, \frac{g(n,e_0)}{\left(n/2\right)^4}u^{-2/3} \Bigg]\,, 
\label{ready_SNR_transition}
\end{eqnarray}

\noindent where formally $N_{\max} \rightarrow \infty$, but in practice, we find that $N_{\max} = 1500$ suffices for all of the eccentricities considered in this work. In Figure~\ref{single_snr}  we show the expected SNR \(\rho\)  for sources that emit in three different regimes:  very low frequencies (\(f_{\rm orb}\ll f_{\rm trans}\)), transition frequencies (\(0.0\,1 f_{\rm trans} \lesssim f_{\rm orb} < f_{\rm trans}\)), and high frequencies (\(f_{\rm orb}\gtrsim f_{\rm trans}\)). These results indicate that:

\begin{itemize}

\item Single resolvable binaries that satisfy \( f_{\rm orb}\ll f_{\rm trans}\) undergo a substantial SNR increase. Heuristically, we can understand this effect based on the results reported in~\cite{peters}, namely, the SNR gets contributions from all harmonics of the orbital frequency \(n\,f_{\rm orb}\), some of which will be located in the region of maximum sensitivity of the PTA. The bottom panel of Figure 5 shows that we can analytically compute the SNR for binaries with orbital frequencies up to \(f_{\rm orb} \lesssim 0.01 f_{\rm trans}\) and \(e_0 \leq 0.8\) with an accuracy better than \(1\%\) using Eqs.~\eqref{first_odd_low_fre},~\eqref{ready_SNR} and~\eqref{prefact} for \(u\ll1\). The bottom panel of Figure~\ref{single_snr} shows that the regime of applicability of these relations increases as we consider sources radiating at very low frequencies (see the line labelled \(u=0.005\)). Using these relations, we find that the increase in SNR in the low frequency regime is given by:

\beq
\rho^{u\leq0.01}_{\rm increase}\equiv \frac{\rho_{e_0\geq0}}{\rho_{e_0=0}} =\sqrt{G(e_0)}\,.
\label{gain_in_snr}
\eeq

Evidently, the contribution from harmonics located in the high frequency regime --- where the sensitivity of the PTA is poorer --- tends to slow down the increase in the SNR and eventually attenuate it. This is clearly shown in the top panel of Figure 5.

\item  Binaries with orbital frequencies \(0.01\, f_{\rm trans} \lesssim f_{\rm orb} < f_{\rm trans}\) need to be described by Eq.~\eqref{ready_SNR_transition} including the contribution from harmonics located in the low and high sensitivity regime frequency of a PTA. In that case we include up to \(N_{\rm max} =1500\) to provide a reliable answer.

\item Finally, binaries with \(f_{\rm orb}\geq f_{\rm trans}\) are very well described by Eqs.~\eqref{second_odd_low_fre},~\eqref{ready_SNR} and~\eqref{prefact} for \(u\geq1\). These relations indicate that the loss in SNR due to eccentricity is given by:

\beq
\rho^{u\geq1}_{\rm loss} \equiv \frac{\rho_{e_0\geq0}}{\rho_{e_0=0}}=\sqrt{Y(e_0)}\,.
\label{loss_in_snr}
\eeq

\end{itemize}

\subsection{Signal-to-noise ratio calculations for a stochastic gravitational wave background}
\label{sec:stochastic}
The nature of a stochastic GW background allows us only to predict the statistical properties of the signal it generates, not the precise signal. Matched filtering approaches are not, therefore, applicable and instead we rely on cross-correlation of data streams from different pulsars. The SNR statistic we shall adopt in this case is described in Ref.~\cite{Moore:2014}. This is the linear combination of cross-correlations between different pulsars that maximizes the SNR, defined as the ratio of the expectation value of the statistic in the presence of a signal to the rms value in the absence of a signal. The SNR for this optimal statistic is
\begin{equation}
\Sigma^2 =8\sum_{i>j}^{N_p}\sum_j^{N_p}\,T_{\rm obs}\int {\rm{d}}f \frac{\Gamma_{ij}^2 S^2_h(f)}{S^2_n(f)}\,.
\label{snr_sto}
\end{equation}
\noindent For an isotropic background, the overlap reduction function \( \Gamma_{ij}\) is entirely determined by the angular separation of the pulsars~\cite{Hellings:1983ApJ}. Assuming that the pulsars in the PTA are randomly placed on the sky, \( \Gamma_{ij}\) can be approximated as the rms value over the sky, i.e.,
\begin{eqnarray}
\label{gamma_cst}
\Gamma_{ij}=\chi&=& \left(4\sqrt{3}\right)^{-1}\,, \\
\label{easy_sum}
\sum_{i>j}^{N_p}\sum_j^{N_p} \Gamma_{ij} &\approx& \frac{N_p\left(N_p-1\right)\chi}{2}\,.
\end{eqnarray}
\noindent Eq.~\eqref{snr_sto} thus takes the form
\begin{equation}
\label{snr_total_sto}
\Sigma^2  = \frac{N_p\left(N_p-1\right)T_{\rm obs}}{12}\int {\rm{d}}f \frac{S^2_h(f)}{S^2_n(f)}\,.
\end{equation}
\noindent Additionally, 
\begin{eqnarray}
S_h(f) &=& \frac{3H^2_0}{2\pi^2} \frac{\Omega(f)_{\rm GW}}{f^3} \quad {\rm and}\nonumber\\ 
 S_n(f) &=& 24\pi^2 \Delta t\, \sigma^2_{\rm rms} f^2\,.
\label{strains}
\end{eqnarray}

\subsubsection{SNR calculations for binaries with fixed eccentricity}
\label{easy_first}
The SNR for a population of binaries with fixed orbital eccentricity can be derived using Eqs.~\eqref{endes}, \eqref{bfunc} and~\eqref{snr_total_sto}:  
\begin{equation}
\label{stochastic_snr}
\Sigma^2  = \frac{N^2_0\,{\cal M}^{10/3}}{3888\,\pi^{14/3}}\frac{N_p\left(N_p-1\right)\,T_{\rm obs}\,B^2(e_0)}{\left(\Delta t\, \sigma^2_{\rm rms}\right)^2}\int_{\hat{f}_0}^{\infty} \frac{{\rm{d}}f}{f^{26/3}}\,.
\end{equation}
\noindent Using the coordinate transformation \(v= f/\hat{f}_0\),  with \(\hat{f}_0 = T^{-1}_{\rm obs}\), we obtain:
\begin{equation}
\label{stochastic_snr_final}
\Sigma^2  =  \frac{N^2_0\,{\cal M}^{10/3}}{29808\,\pi^{14/3}}\frac{N_p\left(N_p-1\right)\,T^{26/3}_{\rm obs}\,B^2(e_0)}{\left(\Delta t\, \sigma^2_{\rm rms}\right)^2} \,.
\end{equation}
\noindent We thus obtain an  expression for the SNR of a stochastic GW background of identical constant eccentricities \(e_0\):

\begin{eqnarray}
\label{tot_snr_ready_sto}
\Sigma^2  &\equiv&  23.49\,B^2(e_0)\, N_p\left(N_p-1\right)\left(\frac{{\cal M}}{10^8 \, \Msun}\right)^{10/3}   \nonumber\\&\times& \left(\frac{T_{\rm obs}}{10\,{\rm yr}}\right)^{26/3}\left( \frac{N_0}{10^{-3} \, {\rm Mpc}^{-3}}\right)^2 \left(\frac{100\,{\rm ns}}{\sigma_{\rm rms} }\right)^4  \nonumber\\&\times&\left(\frac{0.05\,{\rm yr}}{\Delta t}\right)^2\,.
\end{eqnarray}

\begin{figure}[ht]
\centerline{
\includegraphics[height=0.35\textwidth,  clip]{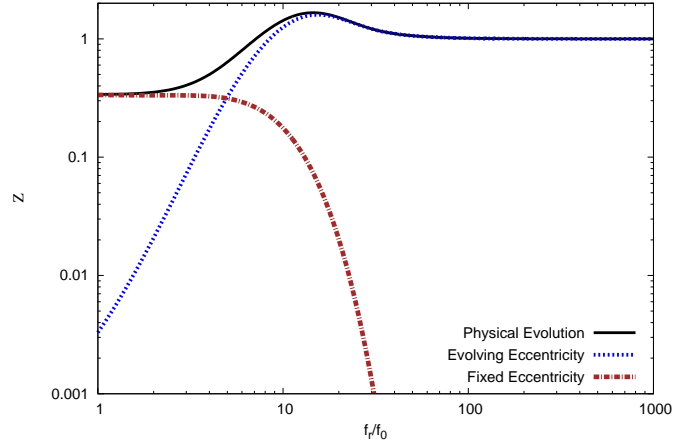}
}
\caption{We show the construction of a function that captures the harmonic content from fixed eccentricity sources with  \(f_{\rm orb}\lesssim f_0\), and that incorporates the contribution from evolving eccentricity sources with \(f_{\rm orb}\gtrsim f_0\). The plot shows the case for \(e_0=0.7\). Notice that the `Physical Evolution' function \(Z\) given by  Eq.~\eqref{physical_s_eq} reproduces the expected physical behavior in the appropriate limits. }
\label{phys_ev}
\end{figure}

\begin{figure}[ht]
\centerline{
\includegraphics[height=0.35\textwidth,  clip]{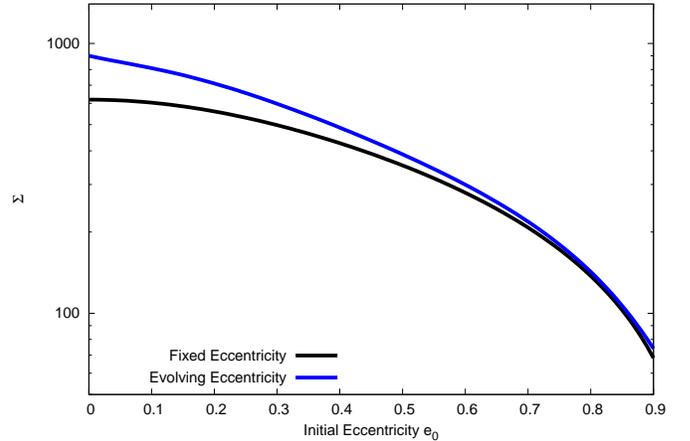}
}
\caption{Expected signal-to-noise ratio \(\Sigma\) for a stochastic gravitational wave background generated by sources with total mass \(M=10^{9}\Msun\) and whose eccentricity is either fixed or evolving, as indicated in the Figure. We have used the fiducial values quoted in the parentheses of Eq.~\eqref{tot_snr_ready_sto}, and assumed \(N_p=10\). Note the substantial suppression in signal-to-noise ratio due to the effect of eccentricity.  As in the case of single resolvable sources, eccentricity noticeably suppresses the detectability when \(e_0\gtrsim 0.6\).}
\label{stochastic_gw_bg_snr}
\end{figure}

\noindent In Figure~\ref{stochastic_gw_bg_snr} we show the expected SNR from a stochastic GW background generated by sources with fixed total mass \(M\). These results have been generated using the fiducial values quoted in parentheses in Eq.~\eqref{tot_snr_ready_sto}, and assuming a network of  \(N_p=10\) pulsars. Figure~\ref{stochastic_gw_bg_snr} shows that eccentricity tends to reduce the expected SNR from a population of compact binary sources. This effect is marginal for binaries with low to moderate values of eccentricity, i.e., for \(e_0\in[0,\,0.6]\). However, the expected SNR of a stochastic GW background generated by a population of highly eccentric binaries satisfies \(\Sigma(e_0=0)\gtrsim 10\,\Sigma(e_0\sim 0.9)\). This is a natural consequence of the effect of the attenuation factor \(B(e_0)\) on the strain of a stochastic GW background (see Figure~\ref{becorr}). In the following Section we extend this analysis to consider populations in which the orbital eccentricity of the binaries evolves.

\subsubsection{SNR calculations for binaries with evolving eccentricity}

A more realistic astrophysical scenario is one in which the eccentricity of binaries that generate the stochastic GW background  is allowed to evolve. In this case, we again use Eqs.~\eqref{snr_total_sto} and~\eqref{strains}, but we now use Eq.~\eqref{endes} along with the function \(S(f, f_0, e_0, z)\) in Eq.~\eqref{l_sum} to take into account the frequency evolution of the eccentricity. Since the function \(S(f, f_0, e_0, z)\)  was derived using Eq.~\eqref{rad_reac}, we identify \(f_0\) as the orbital frequency at which the ensemble of binaries have a fiducial orbital eccentricity \(e_0 = e(f_{\rm orb}= f_0)\), where \(f_0=T^{-1}_{\rm obs}\).

We can compute the SNR for the evolving eccentricity case assuming that the GW background signal evolves both above and below \(f_{\rm orb} = f_0\), so that \(e>e_0\) for \(f_{\rm orb} \lesssim f_0\).  In that scenario, the contribution from sources for frequencies \(f_{\rm orb}\lesssim f_0\) is highly attenuated, as shown in Fig.~\ref{s_function}. However, this scenario is problematic, particularly for large values of \(e_0\). In reality, we expect some dynamical process to be driving binaries to eccentricities of $e_0$ at $f_0$, so that the behavior of the eccentricity at lower frequencies will vary depending on the details of the mechanism. In order to make a SNR comparison between sources with fixed and evolving eccentricity that does not include such severe attenuation for \(f_{\rm orb}\lesssim f_0\), since such attenuation is not well astrophysically motivated, we can modify the framework described by Eq.~\eqref{l_sum}. As discussed by Kocsis and Sesana~\cite{KocSas:2011}, the rate of inspiral depends on the mechanism driving the evolution, and will generically be more rapid than the GW-driven case.  However, given that the likely dynamical processes preceding GW domination tend to drive binary eccentricities to fixed values, one physically reasonable, if simplistic, approach is to assume that sources with \(f_{\rm orb}< f_0\) evolve in frequency at the appropriate rate for gravitational emission, but with constant eccentricity, whereas sources with  \(f_{\rm orb}\gtrsim f_0\) evolve following the behavior given by the function \(S(f, f_0, e_0, z)\) in Eq.~\eqref{l_sum}. Therefore, the attenuation function for this scenario is given by:

\begin{eqnarray}
Z(f, f_0, e_0, z)&=&\sum_{n=1}^{f_r/f_0}\frac{1}{F(e(f_{\rm orb};\, e_0))}\frac{g(n,e(f_{\rm orb};\, e_0))}{\left(n/2\right)^{2/3}} \nonumber\\&+& \sum_{n=f_r/f_0 +1}^\infty\frac{1}{F(e_0)}\frac{g(n,e_0)}{\left(n/2\right)^{2/3}}\,.
\label{physical_s_eq}
\end{eqnarray}

\noindent We show the form of this modified prescription in Fig.~\eqref{phys_ev} assuming a population of sources with eccentricity \(e_0 = e(f_{\rm orb}= f_0)=0.7\). Using this approach, Fig.~\eqref{stochastic_gw_bg_snr} shows that the expected SNR from sources with evolving eccentricity is less attenuated that their fixed eccentricity counterparts, which is a natural consequence of the way in which we constructed the \(Z(f, f_0, e_0, z)\) function, and is the expected physical behavior; since we have found that higher eccentricities are more attenuated, the evolving eccentricity case, which evolves to lower eccentricities due to gravitational-wave emission, should therefore be less attenuated than its fixed eccentricity counterpart. Furthermore, evolving eccentricity sources with low eccentricities  tend to have larger SNR values because \(Z(f, f_0, e_0, z)\gtrsim 1\) for frequencies \(f_r/f_0\lesssim 10\), and most of the SNR is accumulated at lower frequencies due to the strong suppression factor \(f^{-26/3}\) in Eq.~\eqref{stochastic_snr}. Similarly, since highly eccentric systems tend to circularize for larger  \(f_r/f_0\) values, the net enhancement in SNR of evolving over fixed eccentricity sources is less pronounced.

This analysis shows that eccentricity introduces substantial qualitative and quantitative changes in the properties of the GWs emitted that, in the context of current data analysis algorithms, will make their detection more challenging. Developing alternative techniques for the detection and characterization of these signals goes beyond mere curiosity. Since the orbits of SMBH binaries may only shrink to small enough separations for GW domination due to interaction with their environments, and these interactions may drive the binaries to large eccentricity, it is quite plausible that eccentricity will play a fundamental role in the dynamical evolution of SMBH binaries within the sensitivity band of PTAs. This article is a first step to addressing some of these outstanding challenges in the detection of eccentric supermassive binaries.

\section{Conclusions}
\label{sec:conclusions}

Eccentric binary systems may play a more relevant role in the dynamics of compact binary systems than previously thought. In light of studies which suggest that SMBH binaries may have non negligible eccentricity while emitting in the sensitive frequency band of PTAs, we have provided a solid foundation to study the properties of eccentric binary systems.

In this article we have developed an analytical framework that enables the construction of rapid spectra for a stochastic GW background generated by a population of eccentric sources which builds upon the work of Phinney~\cite{Phinney:2001} and Enoki and Nagashima~\cite{Eno:2007}. We have also derived several new analytical approximations that expand upon the results of Peters and Mathews~\cite{peters} to fully assess the impact of including eccentricity on the detection and characterization of eccentric binary systems in the context of single resolvable sources. 

The analytical summations we have derived to benchmark the SNR of single binaries that radiate in the high frequency regime of PTA sensitivity to continuous wave sources do not suffer from the limitations of numerical summation, particularly for very large eccentricities where harmonics at hundreds or thousands of times the orbital frequency may significantly contribute to the signal. Regarding single resolvable binaries that radiate predominantly in the low frequency PTA sensitivity band, our analytical results can be used to benchmark the increase in SNR for sources with eccentricities as high as \(e_0\sim 0.8\) with an accuracy better than \(1\%\).

We have provided ready to use expressions to compute the SNR for eccentric single resolvable sources and a stochastic GW background generated by a population of eccentric binaries. Our results conclusively show that eccentricity will have a positive impact on the detection of single resolvable sources emitting primarily at gravitational-wave frequencies \(f < 2\, T^{-1}_{\rm obs}\). On the other hand, single resolvable sources whose fundamental $n=1$ harmonic is located at a frequency \(f=f_{\rm orb} \geq 2\, T^{-1}_{\rm obs}\), or a stochastic, isotropic GW background generated by binaries with low to moderate values of eccentricity (\(e_0\in[0,\,0.6]\)) may still be recovered with SNRs comparable to their quasi-circular counterparts. The SNRs of highly eccentric binaries, however, will be substantially suppressed, thus requiring the development of alternative search techniques to detect and characterize these signals.

In forthcoming work, we will apply the tools developed here to devise a new, efficient and accurate framework to explore the ability of PTAs to extract the signatures of eccentric binary systems and reconstruct the intrinsic parameters of single resolvable sources and the astrophysical distribution of parameters for stochastic signals.

\section*{Acknowledgments}
JG's work is supported by the Royal Society. This research was in part supported by ST's appointment to the NASA Postdoctoral Program at the Jet Propulsion Laboratory, administered by Oak Ridge Associated Universities through a contract with NASA. We thank Joe Romano for kindly sharing with us his personal notes on the targeted sensitivity of PTAs, which formed the basis of Ref.~\cite{Thrane:2013}, and Chris Moore for verifying that the dimensionless effective noise amplitude for continuous wave sources scales as \(\sim f^{-2}\) in the low frequency regime. This work was supported in part by National Science Foundation 
Grant No. PHYS-1066293 and the hospitality of the Aspen Center for Physics.
\clearpage
\appendix 

\section{Sums of Bessel functions that are relevant for the study of eccentric binary systems}
\label{Apen_PN}
In this Appendix we show how to evaluate the sum over all harmonics \(n\) for the cases described in the main text of the article. The solutions presented in this Appendix are based on Bessel's solution of the Kepler equation,  \(M= e -e \sin E(M,e)\)~\cite{Bessel}:

\begin{equation}
E(M,e) = M + 2\sum_{n=1}^{\infty} \frac{\sin (nM)}{n}J_n (ne)\,.
\label{emeq}
\end{equation}

\noindent Using the previous relation, we have found the following results:

\begin{eqnarray}
&&\sum_{n=1}^{\infty} n^8 J^2_n (ne)= \frac{e^2}{4\left(1-e^2\right)^{25/2}}\Bigg[1 + \frac{973}{4}e^2 \nonumber\\&+&  \frac{40065}{8}e^4 +\frac{1515705}{64} e^6 + \frac{4317789}{128}e^{8} + \frac{7679931}{512}e^{10} \nonumber\\&+& \frac{1779939}{1024}e^{12} +  \frac{385875}{16384}e^{14} \Bigg]\,,\\
&&\sum_{n=1}^{\infty} n^8 J'^2_n (ne)= \frac{1}{4\left(1-e^2\right)^{23/2}}\Bigg[1 + \frac{975}{4}e^2 \nonumber\\&+&  \frac{40701}{8}e^4 +\frac{1585023}{64} e^6 + \frac{4716117}{128}e^{8} + \frac{8832369}{512}e^{10} \nonumber\\&+& \frac{2163231}{1024}e^{12} +  \frac{496125}{16384}e^{14} \Bigg]\,,\\
&&\sum_{n=1}^{\infty} n^7 J_n (ne) J'_n (ne)= \frac{e}{4\left(1-e^2\right)^{21/2}}\Bigg[1 + 117e^2 \nonumber\\&+&  \frac{10809}{4}e^4 +\frac{14091}{4} e^6 + \frac{317205}{128}e^{8} +   \frac{53235}{128}e^{10} \nonumber\\&+&  \frac{7875}{1024}e^{12}  \Bigg]\,,\\
&&\sum_{n=1}^{\infty} n^6 J^2_n (ne)= \frac{e^2}{4\left(1-e^2\right)^{19/2}}\Bigg[1 + \frac{217}{4}e^2 \nonumber\\&+&  \frac{1259}{4}e^4 +\frac{11815}{32} e^6 + \frac{11455}{128}e^{8} + \frac{1125}{512}e^{10}\Bigg]\,,\\
&&\sum_{n=1}^{\infty} n^6 J'^2_n (ne)= \frac{1}{4\left(1-e^2\right)^{17/2}}\Bigg[1 + \frac{219}{4}e^2 \nonumber\\&+&  \frac{1327}{4}e^4 +\frac{13585}{32} e^6 + \frac{14535}{128}e^{8} + \frac{1575}{512}e^{10}\Bigg]\,,\\
&&\sum_{n=1}^{\infty} n^5 J_n (ne) J'_n (ne)= \frac{e}{4\left(1-e^2\right)^{15/2}}\Bigg[1 + 24e^2 \nonumber\\&+&  \frac{255}{4}e^4 +\frac{55}{2} e^6 + \frac{135}{128}e^{8} \Bigg]\,,
\end{eqnarray}

\begin{eqnarray}
&&\sum_{n=1}^{\infty} J^2_n (ne)= -\frac{1}{2} + \frac{1}{2\left(1-e^2\right)^{1/2}}\,,\\
&&\sum_{n=1}^{\infty} n\, J_n (ne)\, J'_n (ne)= \frac{e}{4 \left(1-e^2\right)^{3/2}}\,,\\
&&\sum_{n=1}^{\infty} \left(\frac{J_n (ne)}{n}\right)^2= \frac{e^2}{4}\,,\\
&&\sum_{n=1}^{\infty} \frac{J_n  (ne) J'_n  (ne) }{n}= \frac{e}{4}\,,\\
&&\sum_{n=1}^{\infty} \left(\frac{J'_n (ne)}{n}\right)^2= \frac{1}{4}-\frac{1}{8}e^2\,.
\end{eqnarray}
\noindent Using these results and those quoted in the Appendix of PM \cite{peters}, we obtain
\begin{eqnarray}
\label{gnssq}
L(e) &=& \sum_{n=1}^{\infty}n^4 g(n,\,e) = \frac{16}{\left(1-e^2\right)^{19/2}}\Bigg[ 1 + \frac{16579}{384}e^2 \nonumber\\&+&  \frac{459595}{1536}e^4 +\frac{847853}{1536} e^6 + \frac{3672745}{12288}e^{8} + \frac{1997845}{49152}e^{10} \nonumber\\&+&  \frac{41325}{65536}e^{12} \Bigg]\,,\\
\label{gnsq}
G(e) &=& \sum_{n=1}^{\infty}n^2 g(n,\,e) = \frac{4}{\left(1-e^2\right)^{13/2}}\Bigg[ 1 + \frac{85}{6}e^2 \nonumber\\&+&  \frac{5171}{192}e^4 +\frac{1751}{192} e^6 + \frac{297}{1024}e^{8}\Bigg]\,,\\
\label{we_all_know_you}
F(e)&=&\sum_{n=1}^{\infty}  g(n,e) =\frac{1 + \frac{73}{24}e^2+ \frac{37}{96}e^4}{\left(1-e^2\right)^{7/2}}\,,\\
\label{gns}
H(e) &=& \sum_{n=1}^{\infty}\frac{g(n,\,e)}{n^2} = \frac{4-\sqrt{1-e^2} }{12\sqrt{1-e^2}}\,,\\
\label{gnp4}
Y(e) &=&  \sum_{n=1}^{\infty}\frac{g(n,\,e)}{n^4} = \frac{1}{16} - \frac{e^2}{48}\,.
\end{eqnarray}

\noindent Eq.~\eqref{gns} was used to derive Eqs.~\eqref{pow_rms}-\eqref{sum_rms}. Eq.~\eqref{we_all_know_you} was used in Eq.~\eqref{eval_g}. The remaining expressions,  \eqref{gnsq} and~\eqref{gnp4},  were used to determine  Eqs.~\eqref{gn23}, ~\eqref{first_odd_low_fre} and~\eqref{second_odd_low_fre}.

\section{Convergence of infinite sums}
\label{convergence}

We now estimate how many terms \(n\) are needed for convergence of sums of the type:
\begin{equation}
N(n_{\rm max}) = \sum_{n=1}^{n_{\rm max}}n^{p} g(n,\,e)\,.
\label{num_sol}
\end{equation}
\noindent We do this by computing the fractional error in the numerical value of the sum \(N\) by including up to \(n_{\rm max}\) harmonics, and then comparing this value with the exact analytical result  and the numerical fit. We consider first the well known sum given by Eq.~\eqref{we_all_know_you}. We have found that including up to 100 harmonics is sufficient to reproduce the exact analytical result for eccentricities  \(e\lesssim 0.7\). However, for eccentricities up to  \( e=0.9\) we need to include up to \(n=400\)  harmonics; \(n=800\) for eccentricities up to \(e= 0.94\) and  \(n=1200\) for eccentricities as high as \(e=0.96\).

\begin{figure}[ht]
\centerline{
\includegraphics[height=0.35\textwidth,  clip]{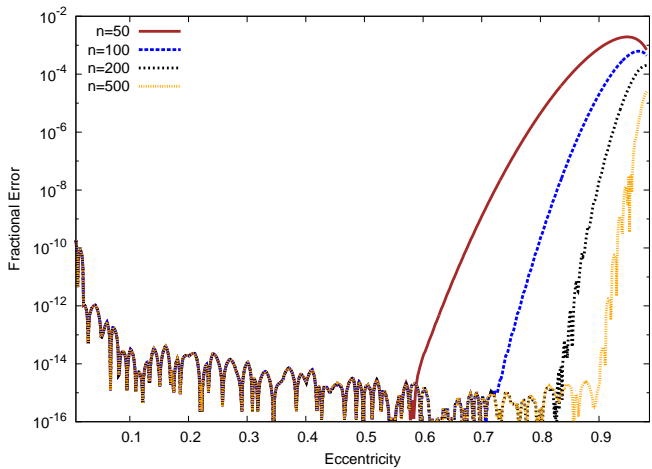}
}
\caption{Fractional error in Eq.~\eqref{num_sol} for \(p=-4\). The choice \(n_{\rm max}=100\) is sufficient to ensure that the \({\rm fractional\, error}\lesssim 0.1\%\) in the entire domain \(e\in[0.0,\,0.95]\).}
\label{my_conv}
\end{figure}

Another important sum is given by  Eq.~\eqref{gnp4}.  Figure~\ref{my_conv} shows that this sum is highly convergent. Note that \(n_{\rm max}=100\) harmonics is sufficient to ensure that the \({\rm fractional\, error}\lesssim 0.1\%\) in the entire domain \(e\in[0.0,\,0.98]\). With \(n_{\rm max}=500\) harmonics, the \({\rm fractional\, error}\lesssim 0.001\%\) in the entire domain \(e\in[0.0,\,0.98]\).

\section{Attenuation factor B(e)}
\label{fundamental_sum}
\begin{figure}[ht]
\centerline{
\includegraphics[height=0.35\textwidth,  clip]{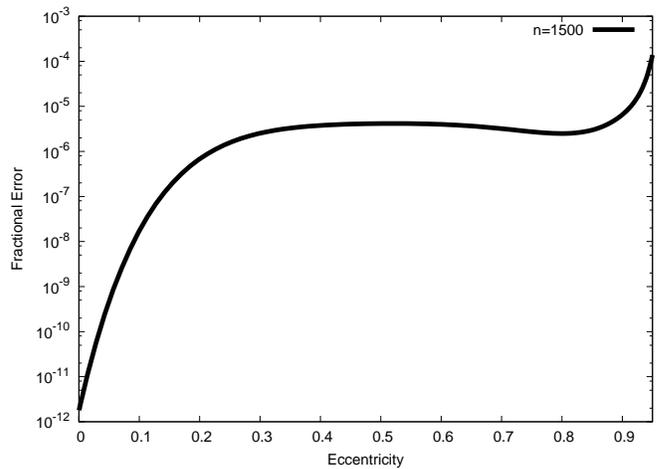}
}
\caption{The numerical fit given by Eq.~\eqref{gn23} reproduces Eq.~\eqref{num_sol} with \(p=-2/3\) and \(n_{\rm max} =1500\) with a \({\rm fractional\, error}\lesssim 0.01\%\)  in the entire domain \(e\in[0.0,\,0.95]\).}
\label{n23_test_convergence}
\end{figure}

We could not find an analytical solution for the fundamental sum given in Eq.~\eqref{gn23}. Instead, we constructed a numerical fit that robustly reproduces the sum given by Eq.~\eqref{num_sol} with \(p=-2/3\) and \(n_{\rm max} =1500\) with a \({\rm fractional\, error}\lesssim 0.01\%\)  in the entire domain \(e\in[0.0,\,0.95]\). This is shown in Figure~\ref{n23_test_convergence}.

\clearpage


\bibliography{references}

\end{document}